\documentclass[10pt,a4paper,twocolumn,english,prb,aps,showpacs,floatfix,groupedaddress,superscriptaddress]{revtex4-1}

\usepackage{graphicx}
\usepackage{epsfig}
\usepackage[english]{babel}
\usepackage{amsmath}
\usepackage{amssymb}
\usepackage{amsfonts}
\usepackage{bm}
\usepackage{bbm}
\usepackage{fixmath}
\usepackage{color,array}
\usepackage{dsfont}

\usepackage[normalem]{ulem}

\newcommand{\neff}{N_\text{eff}}
\newcommand{\nx}[1]{n_\text{#1;max}}
\newcommand{\jq}{J_\text{Q}}
\newcommand{\jqm}{J_\text{Q}^{-1}}
\newcommand{\tr}{\text{Tr}}
\newcommand{\be}{\begin{equation}}
\newcommand{\ee}{\end{equation}}
\newcommand{\bs}{\begin{subequations}}
\newcommand{\es}{\end{subequations}}
\newcommand{\Li}{\mathcal{L}}
\newcommand{\lgl}{\left\langle\left\langle}
\newcommand{\rgl}{\right\rangle\right\rangle}
\newcommand{\ov}[1]{\vec{#1}}
\newcommand{\ntr}{N_\text{tr}}

\newcommand{\Hh}{\mathcal{H}}
\newcommand{\Heff}{\mathcal{H}_\text{eff}}
\newcommand{\s}[2]{{\sigma}_{#1}^{#2}}
\renewcommand{\P}[2]{{P}_{#1}^{#2}}
\newcommand{\m}[1]{\mathbold{#1}}
\newcommand{\mpi}{\m{\pi}}
\newcommand{\sumin}{\sum_{i=1}^N}
\newcommand{\Tr}[1]{\text{Tr}\left(#1\right)}
\newcommand{\ket}[1]{|#1\rangle\rangle}
\newcommand{\levi}[3]{\epsilon_{#1#2#3}}
\newcommand{\Hn}[2]{H_{n_{#1,#2}}}
\begin{document}
\title{Quantum mechanical treatment of large spin baths}

\author{Robin R\"ohrig}
\email{robin.roehrig@tu-dortmund.de}
\affiliation{Lehrstuhl f\"{u}r Theoretische Physik I, Technische Universit\"at Dortmund, Otto-Hahn Stra\ss{}e 4, 44221 Dortmund, Germany}

\author{Philipp Schering}
\email{philipp.schering@tu-dortmund.de}
\affiliation{Lehrstuhl f\"{u}r Theoretische Physik I, Technische Universit\"at Dortmund, Otto-Hahn Stra\ss{}e 4, 44221 Dortmund, Germany}

\author{Lars B. Gravert}
\email{lars.gravert@tu-dortmund.de}
\affiliation{Lehrstuhl f\"{u}r Theoretische Physik I, Technische Universit\"at Dortmund, Otto-Hahn Stra\ss{}e 4, 44221 Dortmund, Germany}

\author{Benedikt Fauseweh}
\email{b.fauseweh@fkf.mpg.de}
\affiliation{Max-Planck-Institut f\"{u}r Festk\"{o}rperforschung, Heisenbergstra\ss{}e 1, D-70569 Stuttgart, Germany}

\author{G\"otz S.\ Uhrig}
\email{goetz.uhrig@tu-dortmund.de}
\affiliation{Lehrstuhl f\"{u}r Theoretische Physik I, Technische Universit\"at Dortmund, Otto-Hahn Stra\ss{}e 4, 44221 Dortmund, Germany}

\date{\rm\today}

\begin{abstract}
The electronic spin in quantum dots can be described by central
spin models (CSMs) with a very large number $\neff\approx 10^4$ to $10^6$
 of bath spins posing a tremendous challenge to theoretical simulations. 
Here, a fully quantum mechanical theory is developed for the limit 
$\neff\to\infty$ by means of iterated equations of motion (iEoM). 
We find that the CSM can be mapped
to a four-dimensional impurity coupled to a non-interacting 
bosonic bath in this limit.
Remarkably, even for infinite bath the CSM does not become
completely classical.
The data obtained by the proposed iEoM approach is tested successfully 
against data from other, established approaches.
Thus, the iEoM mapping extends the set of theoretical tools which
can be used to understand the spin dynamics in large CSMs.
\end{abstract}

%% PACSes
% 03.65.Yz Decoherence; open systems; quantum statistical methods 
% 78.67.Hc Quantum dots 
% 72.25.Rb Spin relaxation and scattering 
% 03.65.Sq Semiclassical theories and applications 

\maketitle

\section{Introduction}
\label{sec.intro}

Since the proposal to use the electronic spin of excess
electrons or holes in quantum dots  \cite{loss98} for the realization of
quantum bits \cite{niels00} an enormous research activity
has started, both experimentally \cite{hanso07,urbas13,warbu13,chekh13b} 
and theoretically \cite{merku02,schli03,coish09,fisch09,ribei13}.
From the theoretical side, the isotropic central spin model (CSM),
first introduced by Gaudin for its integrability \cite{gaudi76,gaudi83},
has become the canonical starting point although various
additional couplings matter as well such as the dipole-dipole
interaction between the nuclear spins forming the spin 
bath \cite{merku02,schli03}, spin anisotropies 
\cite{fisch08a,teste09,hackm14a} including spin-orbit couplings 
\cite{nowac07,ranci14}, and the quadrupolar couplings of the bath spins 
\cite{chekh12,sinit12,hackm15}. In the present work, we
restrict ourselves to the isotropic CSM without further couplings.

In self-assembled quantum dots,  $\neff \approx 10^5$ 
bath spins are relevant \cite{merku02,schli03,lee05,petro08}
or even  $\neff \approx 10^6$ in electrostatically confined 
quantum dots \cite{petta05}. This enormous number makes the reliable
computation of the central spin dynamics extremely challenging
despite the  integrability of the model \cite{farib13a,farib13b}.
Only a few tens of spins can be treated exactly and this remains
true for most numerical approaches as well such as 
exact diagonalization \cite{schli03,cywin10}, Chebyshev expansion (CE)
\cite{dobro03a,dobro03b,hackm14a}, or a direct evolution
of the density matrices via the Liouvillean \cite{beuge16}. 
Density-matrix renormalization group (DMRG)
can cope with up to about 1000 spins, but it is
restricted to short times \cite{stane13,stane14b,grave16}.
Persisting correlations at infinite times 
can be dealt with by mathematically rigorous bounds \cite{uhrig14a,seife16}.
Techniques based on rate equations or on non-Markovian master equations give access to
large bath sizes, but they are well justified only for sufficiently strong
external fields 
\cite{khaet02,khaet03,coish04,breue04,fisch07,ferra08,coish10,barne12,rudne07a}.
So far, the same holds true for an approach based on
equations of motion \cite{deng06,deng08}. 
Finally, cluster expansion techniques are 
powerful, but restricted by the maximum treatable cluster size.
This restriction implies a time threshold up to which the results
are reliable \cite{witze05,witze06,maze08,yang08a,yang09a,witze12}.

The classical counterpart of the CSM approximates
the quantum mechanical spin dynamics well \cite{chen07,stane14b},
see also the related approach based on time-dependent mean-fields
\cite{alhas06,zhang06}.
This behavior can be justified either by the saddle point approximation
for a large spin bath \cite{chen07} or even simpler by the 
quantum fluctuations of the Overhauser field which are suppressed by 
the limit of infinite spin bath \cite{stane13}. Until recently, however,
even the classical CSM could not be treated for
bath sizes comparable to the experimental ones. The Lanczos approach
or the exponential discretization of the spectral density has provided a 
breakthrough to simulate infinitely large systems up to very large times 
\cite{fause17a}. Classical or semiclassical simulations capture
many experimental observations nicely \cite{hackm14b,jasch17}.

Yet, the quantum mechanical dynamics is not fully captured by the
classical simulation. While the argument for classical properties
of the Overhauser field  is strong, there is no such argument
for the central spin $S=1/2$. Thus, there still remains the open issue
to identify specific quantum mechanical effects and to describe them
quantitatively. The measurement of four-point
correlations provides experimental access to quantities which
depend strongly on the sequence of operators acting on the
central spin and hence on its quantumness \cite{press10,becht16,frohl17}.

For these reasons, the present paper proposes an approach to the 
quantum mechanical CSM valid for large spin baths. 
It is based on the equations of motion for spin operators 
\cite{uhrig09c,kalth17} and 
an expansion in the inverse effective number of bath spins $1/\neff$.

The key finding of our approach is that the isotropic CSM can be mapped to a 
four-dimensional impurity coupled to a non-interacting bosonic bath. This mapping
provides an alternative view on the CSM in the limit of a large bath. In order to
establish this mapping, we exploit the simplifying limit of large
spin baths to compute quantum mechanical traces 
of sums of large numbers of spins \cite{seife16}. In this way, we
 evaluate the central spin
autocorrelation function with and without a magnetic field. In the limit $\neff\to\infty$, 
the traces reduce to Gaussian integrals. The approach is benchmarked against 
data from exact methods, available only for small number of bath spins. Even though 
this is not the optimum regime for the application of the developped 
approach, the agreement found is promising.

The setup of the paper is the following. After this introduction
we introduce the model in Sect.\ \ref{sec:model} and 
derive the advocated approach in Sect.\ \ref{sec:approach}. 
Then, we show in Sect.\ \ref{sec:comparison} how its results
compare with data obtained by established techniques in
order to underline the validity of its derivation.
Finally, the results are summarized and an outlook is given
in Sect.\ \ref{sec:conclusion}.

\section{ Model }
\label{sec:model}

In the present article, we focus on the paradigmatic 
isotropic central spin model comprising a central spin $\vec{S}_0$ 
with $S=1/2$ and bath spins $\vec{S}_i$ with $S=1/2$
\begin{align}
\label{eq:csm}
\Hh = \vec{S}_0 \cdot \sum\limits_{i=1}^{N} J_i \vec{S}_i ,
\end{align}
where the $J_i$ are the hyperfine couplings.
The field $\vec{B}$ composed of all bath spins
\begin{align}
\label{eq:overhauser}
\vec{B} =  \sum\limits_{i=1}^{N} J_i \vec{S}_i 
\end{align}
is called the Overhauser field. For concreteness, 
we will consider the following generic set of hyperfine couplings
\begin{equation}
\label{eq:def_J}
J_i = C\exp(-i\gamma), \quad i\in \mathbb{N},
\end{equation}
which decrease exponentially as function
of the parameter $\gamma$. This is the typical scenario encountered in
electronic quantum dots where the coupling is proportional to the probability
of the electronic wave function at the location of the nuclear spin
\cite{merku02,schli03}. In two dimensions, the above parametrization
results from Gaussian wave functions \cite{fause17a} and has been used
in many previous studies as well \cite{coish04,farib13a,farib13b,seife16}.
Note that \eqref{eq:def_J} describes the couplings for a \emph{single} CSM.
If we want to describe an ensemble of quantum dots, 
the results have to be averaged over the distribution of the different couplings
for different quantum dots.

In real quantum dots, further interactions such as dipole-dipole and quadrupolar couplings of the nuclear spins play a role on very long time scales. Here we restrict ourselves to the isotropic CSM \eqref{eq:csm} for simplicity to establish our new approach which solves this model in the physically relevant limit of an infinite spin bath.

We emphasize that we can treat an infinitely large spin bath
because $i$ is not limited, i.e., the total number of
bath spins $N$ may be set to infinity. But the physically relevant number
is the finite \emph{effective} number of bath spins $\neff$, i.e., the number of
bath spins which are substantially coupled. This number
can be defined via the ratio of the squared sum of all couplings 
and the sum of all squared couplings \cite{merku02,schli03,stane13,stane14b,fause17a}. 
The latter is given by
\begin{equation}
\label{eq:def_jq}
\jq^2 = \sum\limits_{i=1}^{N} J_i^2.
\end{equation}
Note that $\jq$ sets the energy scale of the dynamics
on short time scales, i.e., it is set to unity in the
numerical evaluations below. 
The effective number $\neff$ of bath spins reads \cite{fause17a} 
\begin{equation}
\label{eq:neff}
\neff =2/\gamma + O(\gamma).
\end{equation}
Thus, $\gamma\approx 10^{-5}$ to $10^{-6}$ is an excellent small parameter
suitable to control a perturbative approach systematically.
We stress that the normalization $\jq=1$ also
implies that the overall prefactor $C$ in \eqref{eq:def_J}
is given by $\sqrt{2\gamma}$ in the limit of small $\gamma$, 
i.e., it scales like $1/\sqrt{\neff}$.
This means that the contribution of each individual bath spin alone
is  negligible. Only suitable sums over all of them will
have an impact which is relevant in the limit $\neff\to\infty$.

\section{Derivation of the approach}
\label{sec:approach}

\subsection{General equation of motion of operators}
\label{ssc:eom_general}

We are interested in the dynamical spin-spin correlation function of the CSM. Thus we
start from the Heisenberg equation of motion for an arbitrary operator $A$
\be
\label{eq:eom}
\partial_t A = i[\Hh,A] =i\Li A,
\ee
where we introduced the Liouville operator $\Li$ which acts
as linear mapping on the vector space of operators. 
To make the vector space of operators a Hilbert space
we introduce the scalar product of Frobenius type
\be
\label{eq:def_scalar}
\lgl A|B\rgl := \frac{1}{d} \tr (A^\dag B),
\ee
where $d$ is the dimension of the Hilbert space of states. 
Clearly, this definition requires that the local Hilbert space
is finite so that this definition only works for spins or fermions
on discrete sites \cite{kalth17}. Then it corresponds to
the expectation values of the two operators in the limit of infinite temperature where the system is completely disordered so that any state is equally probable. 

Bosonic degrees of freedom can only be
treated at the price of a truncation of their local Hilbert spaces.
Note, however, that the above definition works also for models
with an infinite number of sites as long as the operators $A$ and $B$
affect only a finite number of sites, i.e., they act on finite subclusters.
This is the situation we are dealing with here.

The key advantage of the definition \eqref{eq:def_scalar} 
in comparison to other choices is that the Liouville operator $\Li$
is self-adjoint with respect to this scalar product \cite{kalth17}
\be
\lgl A|\Li B\rgl = \lgl \Li A|B\rgl.
\ee
Hence, the operator dynamics induced by $\Li$ shows oscillatory behavior
or, possibly complicated, superpositions of oscillations. But no
power law or exponential divergences occur, even if truncated
orthonormal operator bases are used. We emphasize that superpositions of
oscillations are precisely what one expects for quantum mechanical models.

The application of the equation of motion to the CSM suggests
to consider an operator basis made from products of components of 
spin operators at different sites \cite{kalth17}. In principle, this does work
and we implemented it (not shown). But we quickly realized 
that one has to track essentially \emph{all} possible
combinations of operators on \emph{all} sites in order to obtain
reliable results. Hence, the direct application of the equations of motion
quickly becomes impractical. 

This conclusion is corroborated by an analytical argument. Suppose for simplicity
that the bath spins do not move. Then the Overhauser field $\vec{B}=B\vec{n}$
is a static magnetic field of size $B$ pointing in the direction
given by the unit vector $\vec{n}$ about which the central spin $\vec{S}_0$ 
precesses according to \cite{merku02}
\begin{align}
\label{eq:rotate}
\vec{S}_0(t) &= \vec{n} (\vec{n}\cdot\vec{S}_0(0)) + [\vec{S}_0(0)-
(\vec{n}\cdot\vec{S}_0(0)) \vec{n}]\cos(Bt)
\nonumber \\
&- \{\vec{S}_0(0)\times\vec{n} \}\sin(Bt).
\end{align}
Note that the sign in the last term has been corrected. This equation
has been used by Merkulov et al. in order to describe the spin dynamics
in quantum dots by averaging it over a Gaussian distribution of the Overhauser field
\cite{merku02}.

Expression \eqref{eq:rotate} reveals that arbitrary high powers of the modulus $B$ of 
the Overhauser field $\vec{B}$ are required in order to capture the dynamics for 
long times. Since the Overhauser field is the weighted sum over all bath 
spin operators it is implied that products with arbitrarily many factors
of spin operators are important. We will come back to this point later.
The second message of the result \eqref{eq:rotate} is that it is not the
individual bath spin which influences the central spin, but sums of them.

\subsection{Generalized Overhauser fields}
\label{ssc:hierarchy}

The time evolution of the central spin is governed by the couplings to the spin bath. However, as we see from Eq.\ \eqref{eq:def_J} and the subsequent normalization implying
that the prefactor $C\propto\sqrt{\gamma}$ 
the individual coupling $J_i$ scales like $\mathcal{O} (1/\sqrt{\neff})$, making it almost negligibly small for a realistic number of bath spins. Hence, the dynamics of the individual bath
spin is not a promising starting point in the limit $\neff \rightarrow \infty$. We conclude 
that it is not the single coupling which is important, 
but rather weighted sums of all couplings. 

The first and most important sum is the Overhauser field $\vec{B}$ itself, including all couplings in linear order. This was first realized in Ref.\ \onlinecite{merku02}. 
However, if we want to describe the time evolution exactly, higher orders of the 
individual couplings  need to be taken into account. Using the same argument as before, 
the non-linear contributions of the individual couplings $J_i$ vanish in the limit $\neff \rightarrow \infty$. But extensive sums of the non-linear contributions remain finite. 
This observation was first used in the efficient description of the dynamics of
the classical CSM \cite{fause17a}, but also carries over to the quantum mechanical case
as we show in the following.

We adopt the idea of introducing generalized Overhauser fields
where the weight of each spin is given by polynomials $p_j$ of the couplings $J_i$ 
\be
\label{eq:polynomials_of_couplings}
	\ov{P}_j := 2\sumin p_j(J_i)\ov{S}_i, \quad j\in\mathbb{N}.
\ee
Note that $p_j(x)$ are polynomials of degree $j$. 
This definition deviates from the classical one by a factor of 2 in order to 
ensure  orthonormalization, see below. First, we require that
the polynomials are orthonormal with respect to the following scalar product
for real functions
\bs
\begin{align}
	\label{eq:onb1}
	\left(p_j|p_k\right) &:= \sumin p_j(J_i)p_k(J_i) 
	\\
	&= \delta_{jk}.
	\label{eq:onb2}
\end{align}
\es
Thus, the theory of orthogonal polynomials tells us that they
can be reconstructed iteratively following the standard Lanczos procedure
\be
	 x p_j(x)=\beta_j p_{j+1}(x)+\alpha_j p_j(x)+\beta_{j-1}p_{j-1}(x)
	\label{eq:lanczos}
\ee
where $p_0(x)=0$, $p_1(x)=x$
so that $\ov{P}_1:=2\vec{B}$ is the usual Overhauser field up to a factor of 2.
The recursion coefficients $\alpha_j$ and $\beta_j$ result from 
\bs
\begin{align}
		\beta_{j-1}&=\left(x{p}_{j}|p_{j-1}\right)\\
		\alpha_j&:=\left(x{p}_{j}|p_{j}\right)\\
		\beta_j&:=\sqrt{\left|x{p}_{j}-\alpha_jp_{j}-\beta_{j-1}p_{j-1}\right|^2}.
\end{align}
\es
For the exponential couplings in Eq.\ \eqref{eq:def_J}, these  
coefficients are explicitly derived in Ref.\ \onlinecite{fause17a} 
for $\gamma \ll 1$. They read
\begin{subequations}
\label{eq:recurs}
\begin{align}
\alpha_j &= \frac{4j^2}{4j^2-1}\sqrt{\frac{\gamma}{2}}
\\
\beta_j &= \frac{\sqrt{j(j+1)}}{2j+1}\sqrt{\frac{\gamma}{2}} .
\end{align}
\end{subequations}
We use these recursion coefficients to illustrate the first three generalized Overhauser fields 
\bs
\begin{align}
	\vec{P}_1 &= 2 \sqrt{\gamma} \sum_{i=1}^{N} \overline{J}_i \vec{S}_i\\
	\vec{P}_2 &= 2 \sqrt{\gamma} \sum_{i=1}^{N} (3 \overline{J}_i - \sqrt{8}) \overline{J}_i \vec{S}_i	\\
	\vec{P}_3 &= 2 \sqrt{\gamma} \sum_{i=1}^{N} \sqrt{3} (5 \overline{J}^2_i - 6 \sqrt{2} \overline{J}_i + 3) \overline{J}_i \vec{S}_i,
\end{align}
\es
where $\overline{J}_i := J_i / \sqrt{\gamma}$.

Using the notation of the generalized Overhauser fields
the Hamiltonian can be denoted
\be
\label{eq:hamilton2}
\Hh = \frac{1}{2} \vec{S}_0 \cdot \ov{P}_1 . 
\ee
For later use we draw the reader's attention to the fact 
that the coefficients $\alpha_j$ and $\beta_j$ are of order $\sqrt{\gamma}$,
see \eqref{eq:recurs} and the comprehensive discussion in Ref.\ 
\onlinecite{fause17a}.

It is known \cite{petti85,viswa94} that the recursion coefficients 
can be understood as the matrix elements of the real symmetric tridiagonal matrix 
\be
	\m{T}=\begin{pmatrix}
	\alpha_1 & \beta_1 & 0 & 0 & \dots \\
	\beta_1 & \alpha_2 & \beta_2 & 0 & \dots \\
	0 & \beta_2 & \alpha_3 &  \beta_3 & \dots \\
	\vdots & \vdots & \ddots & \ddots & \ddots 
	\end{pmatrix}
	\label{eq:tridiagonal}
\ee
which acts on the vector of orthonormal polynomials
\be
	\m{p}(x)=\left(p_1(x),\,p_2(x),\,...\, p_n(x)\right)^T
\ee
according to 
\begin{align}
x\m{p}(x)=\m{T}\m{p}(x).
\end{align}
This representation is useful because it tells us how to truncate. 
Truncating the matrix dimension of $\m{T}$ to the positive integer $\ntr$ we introduce
an effective truncation scheme for the recursion coefficients. We keep those
which lie within the $\ntr\times\ntr$ dimensional upper left submatrix of $\m{T}$.
This has proven to be a powerful approach in the classical calculations \cite{fause17a}.

We point out that due to the construction from orthogonal polynomials, 
the generalized Overhauser fields are orthonormal with respect to the Frobenius norm
\be
\label{eq:P-ortho}
\lgl P^\alpha_j|P^\beta_k\rgl = \delta_{\alpha\beta}\delta_{jk} .
\ee
This is explicitly derived in Appendix \ref{app:ortho}.

\subsection{Higher powers of the generalized Overhauser fields}
\label{ssc:higherpowers}

In the definition of the generalized Overhauser fields $\ov{P}_j$ we introduced higher
powers of the couplings $J_i$ by means of certain orthogonal polynomials.
The generalized Overhauser fields are the above mentioned suitably weighted
sums over the spin operators. The weights are given by the orthogonal polynomials.
The precise form of these polynomials depends on the set of couplings.
However, each generalized Overhauser field is still  linear in the
spin operators, see Eq.\ \eqref{eq:polynomials_of_couplings}. In case of commuting 
classical vectors this was sufficient \cite{fause17a}. 

For the quantum mechanical dynamics we are studying here we need more, namely
higher powers of the generalized Overhauser fields $P^\alpha_j$.
We show here that the appropriate way to consider higher powers of
the Overhauser fields is to consider Hermite polynomials $H_n$ of them.
We emphasize that Hermite polynomials are again orthogonal polynomials,
but they used here for something different than the $p_j$: the $p_j$
are polynomials in the couplings while the Hermite polynomials 
to be introduced are polynomials in the generalized Overhauser fields. 

In order to motivate that we need higher powers of the Overhauser fields
let us consider $A=S_0^z$ and insert it into the right hand side of the 
equation of motion \eqref{eq:eom} yielding $i[H,S_0^z]=(S_0^y P_1^x- S_0^x P_1^y)/2$.
This expression is linear in the Overhauser field. But if we iterate it as we have to
do to capture the quadratic time dependence, we consider $A=S_0^y P_1^x- S_0^x P_1^y$.
A simple calculation reveals that $[H,A]$ comprises terms such as $(P_1^x)^2$,
$P_1^x P_1^z$, $(P_1^y)^2$, or $P_1^y P_1^z$, i.e., quadratic powers
of the Overhauser fields occur. This is just meant as simple 
motivating example. In computing the long-time dynamics arbitrarily high powers 
will arise.

Therefore, we include such higher powers in the operator basis and thus have to consider
their norm and scalar products with other terms in the operator basis. 
Hence, we are facing the evaluation of traces of the general type
\be
\label{eq:trace}
I=\tr\left(P_{j_1}^{\alpha_1} P_{j_2}^{\alpha_2} P_{j_3}^{\alpha_3} 
\ldots P_{j_{n-1}}^{\alpha_{n-1}} P_{j_n}^{\alpha_n} 
\right)
\ee
where the trace refers to the Hilbert space of the bath spins.
The fundamental observation starts from the  normalization 
$\lgl P^\alpha_j|P^\alpha_j\rgl=1$ which results from a weighted
sum over the trace of $(S_i^\alpha)^2=1/4$. Since $\neff$ spins
are substantially coupled each single spin of them contributes
only $O(1/\neff)$. Hence, the prefactor $p_j(J_i)$ of $S_i^\alpha$
in $P^\alpha_j$ is of order $1/\sqrt{\neff}$, see also Eq.\ 
\eqref{eq:recurs}.

For $I$ to take a non-zero value even in the limit $\neff\to\infty$ 
one has to combine the operators $P_{j_m}^{\alpha_m}$ to \emph{pairs}.
Let us call one choice of combining all $P_{j_m}^{\alpha_m}$
to pairs a `pairing'.
In each pair, the sum over the individual bath spins runs over $\neff$ sites
and each summand contributes in order $4/\neff$ so that the
pair yields a non-vanishing contribution. So, the product of all pairs
in each pairing yields a finite contribution in the limit $\neff\to\infty$.
The total value of $I$ results from all possible pairings.

Let us consider triples (generally $n$-tuples with $n>2$)
of the $P_{j_m}^{\alpha_m}$ instead of pairs. By this we mean,
that the summation over the bath spins $S_i^\alpha$ is done
over  trilinear terms such as $S_i^x S_i^y S_i^z$ where each 
factor is taken from one generalized Overhauser field in the triple.
Then the summation is more restricted so that in total less 
summations can be done. This reduces the contribution
by at least a factor $1/\neff$. For instance, the trace
\be
\label{eq:product}
X:=\frac{1}{d}\tr\left(P_{1}^{x} P_{2}^{y} P_{2}^{z} 
P_{1}^{x} P_{2}^{y} P_{2}^{z}  \right)
\ee
splits into a product of three pairs
\begin{align}
\label{eq:pair}
X_\text{pair} &= \left(\sum_{i=1}^N p_{1}^2(J_i)\tr(S^x_i)^2\right)
\left(\sum_{i=1}^N p_{2}^2(J_i)\tr(S^y_i)^2\right) 
\nonumber\\
& \qquad \cdot \left(\sum_{i=1}^N p_{2}^2(J_i)\tr(S^z_i)^2\right)
\end{align}
and into a product of two triples
\be
\label{eq:triple}
X_\text{triple} = K \left(\sum_{i=1}^N p_{1}(J_i)p_2^2(J_i)\tr(S^x_i
S^y_i S^z_i)\right)^2
\ee
with some combinatorial factor $K$ and similar products of a quadruple and a pair
and a single 6-tuple. Note that the trace in \eqref{eq:product}
refers to the Hilbert space of the total system while the traces in
\eqref{eq:pair} and \eqref{eq:triple} refer to the local Hilbert space
of a single bath spin $\vec{S}_i$. The crucial observation is that
$X_\text{pair}$ is precisely unity due to the orthonomalization
of the polynomials $p_j$ while $X_\text{triple}$ is of order $1/\neff$,
i.e., subleading, because there is one summation less. 
The same holds for any combinations of $n$-tuples with $n>2$
so that only pairings need to be considered in leading order.
In particular, we learn that $n$ has to be even. 
Due to the orthonormalization
each pair $P^\alpha_j,P^\beta_k$ yields the factor 
$\lgl P^\alpha_j|P^\beta_k\rgl=\delta_{\alpha\beta}\delta_{jk}$,
cf.\ Eq.\ \eqref{eq:P-ortho}.

We conclude that the computation of the leading term of $I$ in an 
expansion in $1/\neff$ is straightforward. It can
be simplified even further by observing that the computation
of all pairings is exactly what is done in the evaluation
of expectation values of random variables fulfilling Gaussian
distributions. This is the content of Wick's theorem for classical
fields \cite{wegne00}. 

We arrive at the stunning conclusion that
the computation of the leading order of
the  quantum mechanical traces $I$ amounts to the calculation
of classical expectation values of the Gaussian random variables
$P^\alpha_j$ with the correlations given by the scalar product
\eqref{eq:orthonormal}. This was first observed in Ref.\ 
\onlinecite{seife16}. Corrections are of order $1/\neff$.

We emphasize the important implication that the sequence
of the operators in the trace in \eqref{eq:trace} does not matter.
This is obviously true for the classical calculations.
It does not represent a contradiction to the quantum mechanical
nature of the spins in the bath because it only holds for the
 leading contribution in an expansion in $1/\neff$. 
In Appendix \ref{app:commute}, we verify explicitly that 
the non-vanishing commutators are less relevant with respect 
to an expansion in $1/\neff$.

We summarize that the computation of the Frobenius scalar 
product for functions of the generalized Overhauser field 
is tantamount
to computing these functions with respect to Gaussian weights. 
But we emphasize that we are developing a fully quantum mechanical 
treatment although the traces are computed via classical integrals.
These classical Gaussian integrals are identical
to the quantum mechanical traces  up to corrections
of the order of $1/\neff\approx 10^{-5}$ or even smaller.

In view of the Gaussian weights
 it suggests itself to consider Hermite polynomials to
describe general functions of the Overhauser fields
because they are the orthogonal polynomials for a Gaussian
weight function \cite{abram64,brons08}. We define
the normalized Hermite polynomials
\begin{align}
	H_n(x)=\frac{1}{\sqrt{n!}} 
	(-1)^ne^{\frac{x^2}{2}}\frac{d^n}{dx^n}e^{-\frac{x^2}{2}}
	\label{eqn:hermite-polynom-def}
\end{align} 
where the factor $1/\sqrt{n!}$ is added for notational
convenience, see below, beyond the 
standard definition in mathematical text books.
As pointed out above, these Hermite polynomials are orthonormalized with respect 
to a Gaussian weight function $w(x)=(\sqrt{2\pi})^{-1}\exp(-x^2/2)$, i.e., 
they fulfil
\begin{align}
\frac{1}{\sqrt{2\pi}}\int_{-\infty}^\infty 
H_n(x)H_m(x)e^{-\frac{x^2}{2}}dx=\delta_{nm}.
\end{align}
Furthermore, the relations 
\bs
	\begin{align}
	xH_n(x)&=\sqrt{n}H_{n-1}(x)+\sqrt{n+1}H_{n+1}(x)\quad,
	\label{eq:hermite_argument}
	\\
	\frac{d}{dx}H_n(x)&=\sqrt{n}H_{n-1}(x) 
	\label{eq:hermite_ableitung}
	\end{align}
\es
hold and are very well known to physicists because
they are the eigen functions of the harmonic oscillators.
We will exploit the analogy to harmonic oscillators further
below.

The Hermite polynomials provide a transparent way
to include higher powers of the  generalized Overhauser fields
because in leading order in $1/\neff$ we have
\bs
\begin{align}
& \lgl H_n(\P{j}{\alpha}) \left| H_m(\P{j}{\alpha}) \right.\rgl
=\nonumber\\
& \quad =\frac{1}{d}
\Tr{	\left(H_n(\P{j}{\alpha})\right)^\dagger
	H_m(\P{j}{\alpha}) }
\\&\quad =
\frac{1}{\sqrt{2\pi}}
\int_{-\infty}^\infty 
H_n(P_j^\alpha)
H_m(P_j^\alpha)
e^{-\frac{1}{2} 	\left(P_j^\alpha\right)^2 }
dP_j^\alpha 
\label{eqn:spur-integral-reell}
\\&\quad =
\delta_{m n}
\end{align}	
\es
where $d$ stands for the dimension of the underlying Hilbert space of bath spins.

Next, we define operators which will form the basis in the Hilbert space
of operators. As usual, the operators acting on the
central spin are described by Pauli matrices $\s{m}{}$ with 
$m\in\{0,1,2,3\}$ where $\sigma_0$ is the identity matrix.
Furthermore, we have to describe powers of the generalized
Overhauser fields to describe the dynamics of the spin bath.
As explained above, Hermite polynomials with the generalized Overhauser fields as arguments
are the appropriate choice because they imply orthonormality of the operator basis.
Concretely, we use the shorthand
\be
{\Hn{j}{\alpha}:=\Hn{j}{\alpha}(\P{j}{\alpha})},
\ee 
where $n_{j,\alpha}$ carries its two subscripts because the degree
of the Hermite polynomial depends on the index of the Overhauser field and
its component $\alpha\in\{x,y,z\}$. Then, a general  basis operator reads
\bs
\begin{align}
	\hat{b}^m_{\m{n}}& :=\s{m}{} A_{\m{n}}
		\label{eqn:hermite_basisoperator}
	\\
	A_{\m{n}} &:= \Hn{1}{x}\Hn{1}{y}\Hn{1}{z}\Hn{2}{x}...
	\\
	\m{n} &:=(n_{1,x},\,n_{1,y},\,\ldots,\,n_{\ntr,z}),
	\label{eq:degree-sequence}
\end{align}
\es
where the $3\ntr$-tuple $\m{n}$ as defined above stores the
degrees of the respective Hermite polynomials.
For the notation to be unique, the sequence of
non-negative integers $n_{j,\alpha}$ is defined as 
shown in \eqref{eq:degree-sequence}.
But for the evaluation of traces and hence of the Frobenius
scalar products it does not matter in leading order in $1/\neff$.

The orthonormality of the $\hat{b}_m^{\m{n}}$ results from
\bs
\begin{align}
	&\lgl\hat{b}^k_{\m{n}}\left|\hat{b}^l_{\m{m}}\right.\rgl
	= \nonumber
	\\
	&\quad = 
	\frac{1}{d}
	\Tr{
		\prod_{j=1}^{\ntr}\prod_{\alpha}
		\left(\sigma_k \Hn{j}{\alpha}\right)^\dagger
		\sigma_l H_{m_{j,\alpha}}
	}
	\\&\quad =
	\frac{1}{2}
	\Tr{\sigma_k
		\sigma_l
	}
	\frac{1}{{2^N}}
	\Tr{
		\prod_{j=1}^{\ntr}\prod_{\alpha}
		\Hn{j}{\alpha}
		H_{m_{j,\alpha}}
	}
	\\&\quad =
	\delta_{kl}
	\delta_{\m{n}\m{m}},
	\end{align}	
\es
where we used again that the traces can be computed by Gaussian integrals and
that the Hermite polynomials are orthonormal with respect to these integrals.
The Kronecker symbol $\delta_{\m{n}\m{m}}$ is unity if both sequences
of non-negative integers $\m{n}$ and $\m{m}$ are equal; otherwise it vanishes.
Thus, the basis spanned by the operators $\hat{b}_m^{\m{n}}$ 
provides an excellent starting point to
treat the equations of motion of the CSM quantitatively.

\subsection{Specific equation of motion}
\label{ssc:eom_specific}

In this next step, we establish the equations of motion
for the developed basis of operator. Hence, we have to
know from where we start and how the Liouville operator
acts on $\hat{b}^m_{\m{n}}$.

We aim at the isotropic CSM without magnetic field in the
first place. We intend to compute the $\langle S^z(t) S^z(0)\rangle$ correlation 
as function of time.
Thus, the initial basis operator is the $z$-component of the
central spin while all bath spins are supposed to be in a completely
disordered state, i.e., there is no operator acting
on any bath spin. This choice is indicated by the extremely 
small couplings $J_i$ which are exceeded by the thermal energy
even at 10K so that the spin bath can be considered to be at infinite
temperature \cite{coish09,urbas13}.
Thus, initially the spin bath is described by the Hermite polynomials $H_0$ equal
to the identity for all $j$ and all $\alpha$, i.e.,
 $H_0(\P{j}{\alpha})=\mathds{1}$. The starting operator is
\begin{align}
\hat{b}^3_{\m{o}}=\s{3}{}H_0H_0H_0 ... .
\end{align}

In order to obtain the time evolution we have to compute
the action of $\Li$ on the basis operators which amounts
to computing the commutator between the Hamiltonian $\Hh$ in 
\eqref{eq:hamilton2} and $\hat{b}^m_{\m{n}}$,
which is a product of an operator $\sigma_m$ acting on the
central spin and of the operator $A_\m{n}$ acting
on the spin bath. Similarly, the Hamiltonian consists of a sum of products of
an operator acting on the central spin and an operator 
acting on the spin bath.
If we use $C$ and $C'$ for operators of the central spin and $A$ and $A'$ for 
operators acting on the bath, the structure of the commutator is
\bs
\begin{align}
\label{eq:comm1}
[CA,C'A'] &= [C,C']AA' + C'C[A,A']
\\
\label{eq:comm2}
&= [C,C']A'A + CC'[A,A']
\end{align}
\es
where both right hand sides are equivalent. This appears to pose a 
problem because it introduces an ambiguity. In leading order in 
$1/\neff$  the first terms in Eqs.\ \eqref{eq:comm1} and \eqref{eq:comm2} 
are indistinguishable
because the sequence of $A$ and $A'$ does not matter. In return, only
the average of the second term can matter. Hence, we use the symmetrized
relation
\bs
\begin{align}
[CA,C'A'] &= T_1 +T_2
\\
\label{eq:t1def}
T_1 &= \frac{1}{2} [C,C']\{A,A'\}
\\
\label{eq:t2def}
 T_2 &= \frac{1}{2} \{C',C\}[A,A']
\end{align}
\es
which avoids the ambiguity.

We consider $\Li(\sigma_m A_{\m{n}})$ and attribute the resulting terms
to $T_1$ or to $T_2$ depending on whether they result from
the commutation of two operators of the central spin or from the
commutation of two operators of the spin bath.
The explicit calculation of $T_1$ and $T_2$ is performed in 
Appendix \ref{app:t_1_t_2}. The resulting expression for $T_1$ is
\begin{align}
&T_1[\Li(\sigma_m A_{\m{n}})] =
\nonumber \\
&\quad \frac{i}{4}
\left(\sigma_{m+1}\{\P{1}{m-1},A_\m{n}\} -\sigma_{m-1}\{\P{1}{m+1},A_\m{n}\} 
\right).
\label{eq:t1}
\end{align}
To denote $T_2$ concisely, we need two definitions
\begin{align}
\label{eq:def_R}
R_j^\alpha &:= 
		\beta_j{P}_{j+1}^\alpha+\alpha_j{P}_j^\alpha+\beta_{j-1}{P}_{j-1}^\alpha
\end{align}
and the mapping $\mpi$ with
\be
\label{eq:mpi}
\mpi(\m{n},j,\alpha) := (n_{1,x},\,n_{1,y},\,\ldots,\,n_{j,\alpha}-1,\,
\ldots,\,n_{\ntr,z}),
\ee
which means that $\mpi(\m{n},j,\alpha)$ maps the sequence $\m{n}$ to the same sequence except
that the degree $n_{j,\alpha}$ is decremented by one.
Then $T_2$ reads
\bs
\label{eq:t2}
\be
T_2[\Li(\sigma_m A_{\m{n}})] =  \frac{-i\sigma_0}{2} \sum_{j=1}^{\ntr}
		\sum_{\alpha,\delta=1}^3 \levi{\alpha}{m}{\delta} \sqrt{n_{j,\alpha}}
		R^\delta_j A_{\mpi(\m{n},j,\alpha)}
\ee
for $m\in\{1,2,3\}$. For $m=0$, we obtain
\be
T_2[\Li(\sigma_0 A_{\m{n}})] =  \frac{-i}{2} \sum_{j=1}^{\ntr}
		\sum_{\alpha,\beta,\delta=1}^3\levi{\alpha}{\beta}{\delta} 
		\sqrt{n_{j,\alpha}} \sigma_\beta R^\delta_j	A_{\mpi(\m{n},j,\alpha)}.
\ee
\es

The sum of Eqs.\ \eqref{eq:t1} and \eqref{eq:t2} 
yields the action of $\Li$ on a general basis operator 
$\hat{b}^m_{\m{n}}=\sigma_m A_\m{n}$
concluding the present subsection. However, we will not use
these equations in their present form in explicit calculations 
because they are a bit cumbersome.
Instead, we will use creation and annihilation operators to
denote the action on the Hermite polynomials as is done 
for the harmonic oscillator in any text book on quantum mechanics.
This is introduced in the next subsection.

\subsection{Effective Hamilton operator without external magnetic fields}
\label{ssc:effectiv_ham_without}

We view the Hilbert space of operators as conventional
Hilbert space of states and denote the basis operators
by kets
\bs
\begin{align}
	\ket{\hat{b}^m_{\m{n}}}&=\ket{m;\m{n}}
	\\
	&=\ket{m;n_{1,x}n_{1,y}n_{1,z},n_{2,x}\,...}.
\end{align}
\es
We remind the readers that the scalar product of these operator kets
is given by the trace $\lgl A|B\rgl:=\tr(A^\dag B)/d$.
In order to express the Liouville dynamics in terms of operators
by a Hamiltonian dynamics on operator kets we are looking for
an effective Hamiltonian which fulfills
\begin{align}
	\frac{d}{dt}\ket{m;\m{n}}
	=
	-i\Heff \ket{m;\m{n}}
	\label{eqn:bewegungsgleichung-dirac}.
\end{align}
Once we have found this effective Hamiltonian
$\Heff$ we can use any analytic or numerical
tool developed to deal with Hamiltonian dynamics.

Inspecting Eq.\ \eqref{eq:t2} 
one realizes that a single $H_n$ is transformed to
$\sqrt{n} H_{n-1}$ which is precisely the action 
of a bosonic annihilation operator $a$, known
from the analytic solution of the harmonic oscillator. 
The other occurring action on Hermite polynomials
is the multiplication with their argument, see
Eq.\ \eqref{eq:hermite_argument}. This is represented
by the sum $a+a^\dag$ of the bosonic annihilation and creation
operator. 

Since the Hermite polynomials denoted by $\Hn{j}{\alpha}$
depend on different arguments $\P{j}{\alpha}$,
we have to introduce different bosonic operators depending
on the labels $j,\alpha$. Thus, we use 
$a_{j,\alpha}^{\phantom{\dag}}$ and $a^\dag_{j,\alpha}$.
These operators are sufficient to describe the action
of $\Li$ on the spin bath.

In addition, we have to describe the action of $\Li$ on the
central spin, i.e., on the Pauli matrices describing
the operators acting on the central spin. 
There are two kinds of processes, namely anticommutation and
commutation, see Eqs. \eqref{eq:t1def} and\eqref{eq:t2def}. 

The anticommutation with $\sigma_k$ 
is described by a matrix $M_k$ with $k\in\{1,2,3\}$, i.e.,
its matrix elements are computed by means of the scalar
product
\bs
\label{eq:pauli_anticommut}
\begin{align}
\lgl n|M_k |m\rgl &:= \frac{1}{2}\lgl\sigma_n| \{\sigma_k,\sigma_m\} \rgl
\\
&=\begin{cases}
	\delta_{mk} + \delta_{nk} & \text{if } nm=0
	\\
	0 & \text{otherwise}
\end{cases}
\end{align}
\es
for $n,m\in\{0,1,2,3\}$.
Note there is no need to introduce $M_0$ because it equals the $4\times4$ 
identity matrix. These matrices are given explicitly in Appendix \ref{app:matrices}.

The commutation with $\sigma_k$ 
is described by a matrix $K_k$ with $k\in\{1,2,3\}$, i.e.,
its matrix elements are computed by means of the scalar
product
\bs
\label{eq:pauli_commut}
\begin{align}
\lgl n|K_k|m\rgl &= \frac{1}{2}\lgl\sigma_n| [\sigma_k,\sigma_m]\rgl
\\
&=\begin{cases}
	0 & \text{if } nm=0
	\\
	i\levi{n}{k}{m} & \text{otherwise}
\end{cases}
\end{align}
\es
for $n,m\in\{0,1,2,3\}$. There is no need to introduce $K_0$ because it 
vanishes completely. These matrices are given explicitly in Appendix \ref{app:matrices}.

\begin{figure}[htb]
	\centering
	\includegraphics[width=\columnwidth]{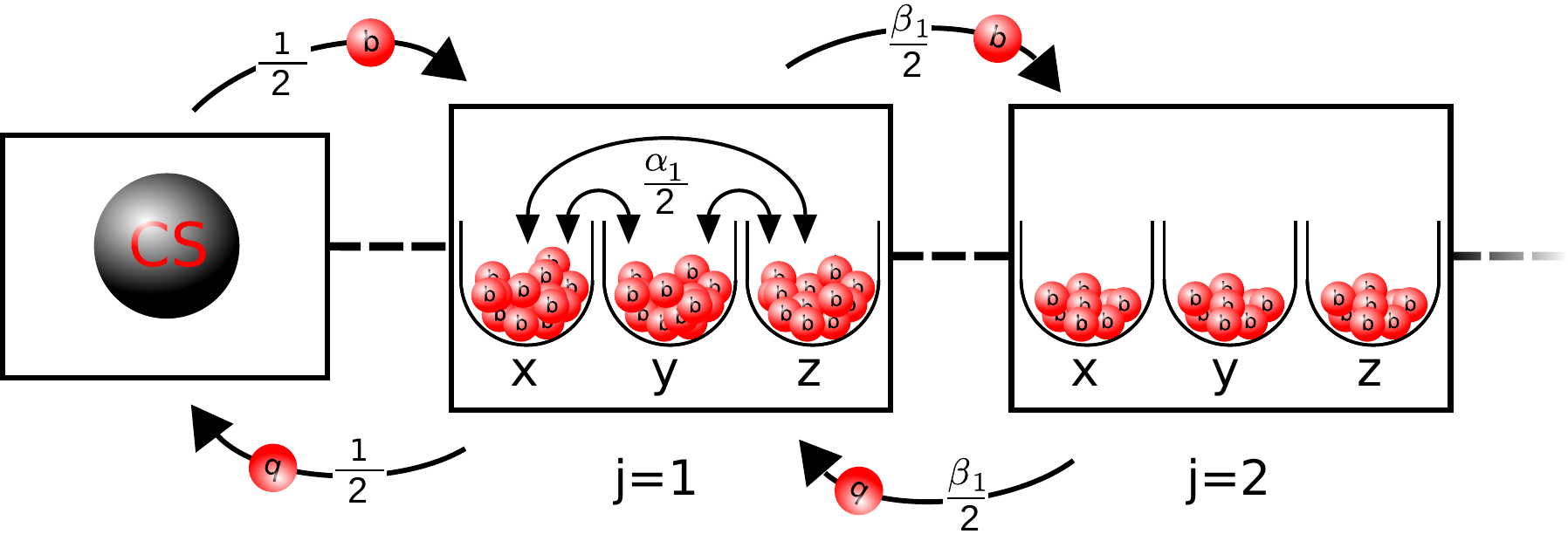}
	\caption{Sketch of the flow of bosons denoted `b' in the
	system described by $\Heff$ comprising the two
	terms given in \eqref{eq:twoterms}. The components
	$x,y,z$ in the figure are denoted by $1,2,3$ in the equations
	for notational simplicity. In addition to what is sketched, 
	each local process proportional
	to $\alpha_j$ or hopping process proportional to $\beta_j$  also has
	an effect on the head of the chain, i.e., on CS, as expressed
	in \eqref{eqn:H_OFb} by the matrices $M_\beta$.
	\label{fig:chain}}
\end{figure}

It is suitable to split the effective Hamiltonian in its parts resulting
from the terms of type $T_1$ and from the terms of type $T_2$, respectively.
Thus, we consider
\begin{align}
\label{eq:effect_hamil}
	\Heff=\Heff^\text{CS} + \Heff^\text{ch},
\end{align}
where the first term $\Heff^\text{CS}$ constitutes the head, i.e., 
site 0,  of a semi-infinite chain of sites $j\in\mathds{N}$, see 
Fig.\ \ref{fig:chain}.
The second term $\Heff^\text{ch}$ describes the action on the chain.
These two terms read
\bs
\label{eq:twoterms}
\begin{align}
\label{eq:HCS}
\Heff^{\text{CS}} &= \frac{1}{2}\sum_{\alpha=1}^3 
{K}_\alpha
\left(
a_{1,\alpha}^{\phantom{\dagger}}+a_{1,\alpha}^\dagger
\right)
\\
\Heff^{\text{ch}} &=\frac{i}{2}\sum_{j=1}^{\ntr}\sum_{\alpha,\beta,\delta=1}^3
\levi{\alpha}{\beta}{\delta}{M}_\beta
a_{j,\alpha}^{\phantom{\dagger}}
\left\{ \alpha_j
(a_{j,\delta}^{\phantom{\dagger}}+a_{j,\delta}^\dagger) \right.
\nonumber
\\&  +
\beta_{j-1}
(a_{j-1,\delta}^{\phantom{\dagger}}+a_{j-1,\delta}^\dagger)
+ \left.\beta_{j}
(a_{j+1,\delta}^{\phantom{\dagger}}+a_{j+1,\delta}^\dagger)
\right\}.
\label{eqn:H_OFb}
\end{align} 
\es

If we expand the above expression for $\Heff^{\text{ch}}$,
bilinear terms in the annihilation operators appear which seem
to violate the hermiticity of the Hamiltonian. But the antisymmetry
of the Levi-Civita tensor $\levi{\alpha}{\beta}{\delta}$ ensures
that all non-hermitian terms cancel out and that
$\Heff^{\text{ch}} = \Heff^{\text{ch},\dag}$ holds
\begin{align}
\Heff^{\text{ch}} &=\frac{i}{2}\sum_{j=1}^{\ntr}\sum_{\alpha,\beta,\delta=1}^3
\levi{\alpha}{\beta}{\delta}{M}_\beta
\left\{ \alpha_j a_{j,\delta}^\dagger a_{j,\alpha}^{\phantom{\dagger}} \right.
\nonumber\\
& \qquad + \left.
\beta_{j} ( a_{j+1,\delta}^\dagger a_{j,\alpha}^{\phantom{\dagger}} -
 a_{j,\alpha}^\dagger a_{j+1,\delta}^{\phantom{\dagger}}
) \right\}.
\label{eq:chain-hamil}
\end{align}
There is even another step towards diagonalization possible
which is presented in Appendix \ref{app:chain-diag}. We do not
use it in the present article, but it will be useful
in future extensions of the iEoM approach introduced here.

This completes the mapping of the equations of motion for 
operators of the CSM with large spin bath onto an effective
Hamiltonian. This effective Hamiltonian acts on a four-dimensional impurity
described by the matrices $M_k$ and $K_k$ and to a semi-infinite
chain of bosons, see Fig.\ \ref{fig:chain}. The bosons act on the polynomials
of the generalized Overhauser fields while the operator space of the 
central spin is represented by the impurity. Thus, it is now possible
to represent every observable as a  ket of this impurity model. 
For instance an operator acting solely on the central spin is a product state
of the impurity state and the boson vacuum. In the course of the time evolution
of the operator, contributions from the generalized Overhauser fields will appear,
which correspond to the creation, annihilation, or the hopping of the bosons.

The first term of $\Heff$, namely $\Heff^{\text{CS}}$,
creates and annihilates bosons at the head of the chain. Its coupling
constant is $\jq=1$, i.e., relatively large.
The second term of $\Heff$, namely $\Heff^{\text{ch}}$, 
does not change the number of bosons, but lets them change flavor 
($x,y,z$ or $1,2,3$, respectively, depending on the notation) on-site
or combined with hops along the chain. Each change of flavor also changes
the state of the impurity at the head of the chain. The scale of 
$\Heff^{\text{ch}}$ is given by $\sqrt{\gamma}\jq$. Hence, the
corresponding rate is two to three orders of magnitude lower
for $\gamma\approx10^{-5}$ to $10^{-6}$ than the one induced by $\Heff^{\text{CS}}$.
This implies that $\Heff^{\text{ch}}$
is a perturbation to $\Heff^{\text{CS}}$, i.e., 
$\Heff^{\text{ch}}$ is responsible for the slow, long-term dynamics.

\subsection{Effective Hamilton operators for external magnetic fields}
\label{ssc:effectiv_ham_magnetic}

In this subsection, we point out how magnetic fields
can be incorporated as well. First, we study the case where 
$h$ is  acting on the central spin along the $\alpha$ direction. 
If the spin Hamiltonian is amended by the Zeeman term
\be
\label{eq:zeeman}
	\Hh^\text{Z} = -hS_0^\alpha
\ee
this translates straightforwardly to the amendment
\be
\label{eq:eff-zeeman}
	\Heff^\text{Z}=-h K_\alpha
\ee
of the effective Hamiltonian 
because only the operators of the central spins are relevant
and the commutation of Pauli matrices is represented by
the matrices $K_\alpha$. The Zeeman term for the central
spin will be employed in the numerical evaluation
below.

Second, we consider the action of a magnetic field in $\alpha$
direction on the spins of the bath. Motivated by the much lower
nuclear magnetic moments, we denote the magnetic
field by $zh$ where $z\approx1/800$ takes the much lower
effect of a magnetic field on the nuclear spins into account \cite{beuge17}.
This means that we deal with the additional term $\Hh^\text{nZ}$
for the nuclear Zeeman effect
\be
\label{eq:nuc_zeeman}
	\Hh^\text{nZ} = -zh \sumin S_i^\alpha.
\ee
Computing the corresponding Liouville operator yields
\begin{align}
	\Li^\text{nZ} \ov{P}_j =
	i	z\ov{h}\times \ov{P}_j,
	\label{eq:liouville_nz}
\end{align}
which translates to 
\begin{align}
\label{eq:kreuzprodukt_nz}
	\Li^\text{nZ} \P{j}{\alpha} =
	i z\sum_{\beta,\delta} \levi{\alpha}{\beta}{\delta}
	h^\beta \P{j}{\delta}.
\end{align}
Using again (as in the previous  subsections) that the sequence
of bath operators does not matter in leading order in
$1/\neff$, we obtain
\begin{align}
	[\Hh^\text{nZ},\Hn{j}{\alpha}] =
	iz\sqrt{n_{j,\alpha}}H_{n_{j,\alpha}-1} 
	\sum_{\beta,\delta} \levi{\alpha}{\beta}{\delta}
	h^\beta \P{j}{\delta}
	\label{eqn:kommutator_hermite_voll_nZ}
\end{align}
and finally
\be
\Li^\text{nZ}(\sigma_m A^{\m{n}}) =  iz\sigma_m \sum_{j=1}^{\ntr}
		\sum_{\alpha,\beta,\delta} \levi{\alpha}{\beta}{\delta} \sqrt{n_{j,\alpha}}
		h^\beta\P{j}{\delta} A_{\mpi(\m{n},j,\alpha)}.
\ee

This allows us to express the  Liouvillean of the nuclear Zeeman effect
by annihilation and creation operators
\bs
\label{eq:nZ}
\begin{align}
\Heff^{\text{nZ}} &= -iz\sum_{j=1}^{\ntr}\sum_{\alpha,\beta,\delta=1}^3
\levi{\alpha}{\beta}{\delta} h^\beta
a_{j,\alpha}^{\phantom{\dagger}}
(a_{j,\delta}^{\phantom{\dagger}}+a_{j,\delta}^\dagger)
\\
&= 
{-iz}\sum_{j=1}^{\ntr}\sum_{\alpha,\beta,\delta=1}^3
\levi{\alpha}{\beta}{\delta} h^\beta
 a_{j,\delta}^\dagger a_{j,\alpha}^{\phantom{\dagger}}.
\end{align} 
\es
In passing to the second line, we used the antisymmetry
of the Levi-Civita symbol.
Note that the Hamiltonian $\Heff^{\text{nZ}}$ has no effect
on the central impurity so that no matrices $M_\beta$
or $K_\beta$ occur. We will not use this term in the numerical
implementation below. For short times, it is not relevant
due to the small value of $z$, but for longer times
it does have an effect on higher correlations \cite{frohl17} and on 
quantum dots subject to pulses \cite{beuge17,jasch17}.

Thus, magnetic fields acting on the central spin or 
on the bath spins can also be accounted for easily.

\section{Comparison to other data}
\label{sec:comparison}

To establish the validity of the derived effective model in the 
limit $\neff\to\infty$, we compare its results
to quantitative results obtained by various
established approaches which capture the temporal dependence up
to a certain time and which can cope only with relatively small
baths. Still, this is sufficient to see that the effective model
reproduces the correct physics in the limit $\neff\to\infty$.

The effective model of a semi-infinite bosonic chain coupled
to a four-dimensional impurity can be simulated
numerically. The spin-spin correlation in the original model
\be
\label{eq:autocorr}
S(t):=\langle S^z_0(0) S_0^z(t)\rangle
\ee
is obtained in the effective model from the time evolution
of the state $\ket{3;\m{0}}$ where $\m{0}=(0,0,0,\ldots)$ stands
for the bosonic vacuum and $3$ stands for the third Pauli matrix,
i.e., a particular state of the four-dimensional central impurity.
Hence, we compute
\begin{align}
S(t) = \frac{1}{4}\lgl 3;\m{0}| \exp(-i\Heff t) | 3;\m{0} \rgl.
\end{align}
Note that $S(t=0)=1/4$ as it has to be for 
the spin-spin autocorrelation in \eqref{eq:autocorr}.

\subsection{Numerical implementation}

For the present proof-of-principle, we do not implement a
highly sophisticated code to compute the time dependence
induced by $\Heff$. We use a finite, truncated basis
comprising the four states of the central impurity and 
a finite number of bosonic states given by the occupation
numbers in the $3\ntr$-tuple $\m{n}$. The resulting finite
dimensional Schr\"odinger equation for the kets 
is an ordinary linear differential equation which 
is solved by a Runge-Kutta algorithm of fourth order.
The starting vector is $\ket{3;\m{0}}$.
Finally, the scalar product is computed with respect
to the bra $\lgl 3;\m{0}\right|\right.$.

The key approximation in the implementation is the 
truncation of the maximum bosonic occupation for each
site $j\le\ntr$. We restrict the
Hilbert space by means of  the condition 
\begin{align}
 \sum^3_{\alpha=1} n_{j,\alpha} <	\nx{j}.
 \label{eq:truncation}
\end{align}
Note that this  restricts the total number of bosons at
each site $j$ of the chain irrespective of their flavor
$\alpha\in\{1,2,3\}$. This has turned out to be the
most efficient way of local truncation. 

The most important number is the
threshold for the bosons at $j=1$ because it restricts how
well the dominant $\Heff^\text{CS}$ is represented. The other thresholds
can be chosen significantly lower, see below.
In practice, we build the basis iteratively by applying
$\Heff$ again and again. States which do not fulfil 
\eqref{eq:truncation} for $j>1$ are truncated and the repeated application 
of $\Heff$ is continued. For $j=1$, the recursive application
of $\Heff$ is stopped once \eqref{eq:truncation} is violated.
This procedure enhances the performance without changing the
results noticeably up to the times for which the
implementation provides reliable data.

\begin{figure}[htb]
	\centering
	\includegraphics[width=\columnwidth,clip]{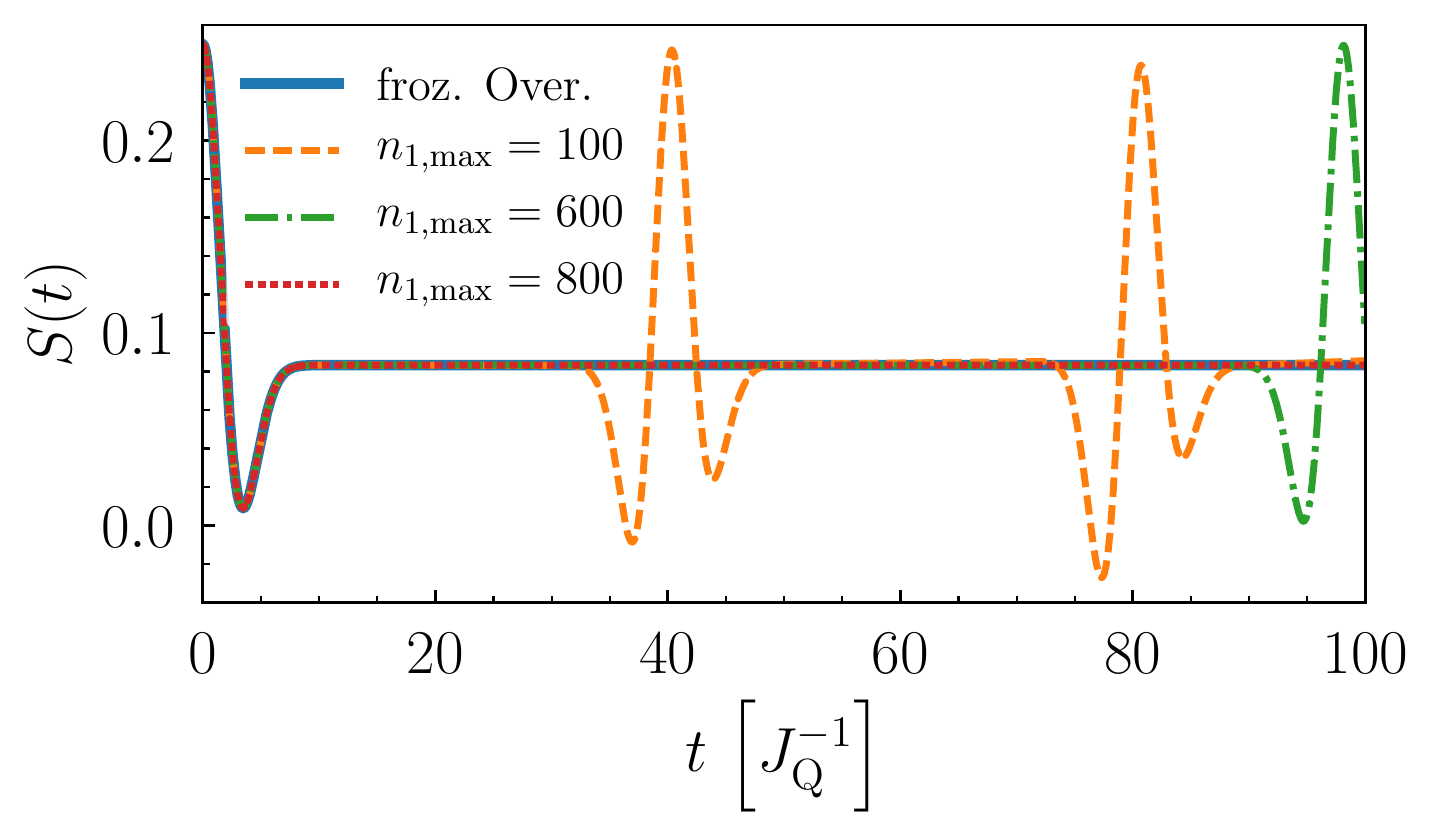}
	\caption{Comparison of the analytical solution $S_\text{fOver}(t)$
	for frozen Overhauser field in Eq.\ \eqref{eq:fover} to the
	numerical result obtained from $\Heff^\text{CS}$ alone
	for various cutoffs $\nx{1}$ of the occupation number.
	\label{fig:fover}}
\end{figure}

First, we consider $\Heff^\text{CS}$ alone so that
only $\nx{1}$ matters. Physically, this means that
the Overhauser field $\ov{B}$ is frozen, i.e., static.
Then, the result can be obtained analytically either
in a fully classical way \cite{merku02} 
or based on correlated fluctuations \cite{stane13} as well. It reads
\be
S_\text{fOver}(t)=
\frac{1}{12}\left\{
1+2\left[1-\frac{\jq^2}{4}t^2
\right] \exp \left[-\frac{\jq^2}{8}
t^2 \right]\right\}
\label{eq:fover}
\ee
and it is  suitable to check our numerical approach on the simplest level. 
In Fig.\ \ref{fig:fover}, the analytical result 
is compared to $S(t)$ obtained from $\Heff^\text{CS}$
alone for various cutoffs $\nx{1}$ in the occupation number.
Clearly, the short-term behavior is perfectly reproduced and
the constant plateau as well up to some threshold time $t_\text{thresh}$
at which a spurious revival of the initial correlation
appears. This spurious revival occurs the later the larger
the cutoff $\nx{1}$ is chosen which underlines that it is
a numerical effect due to the truncation of the Hilbert space.

We analyzed the scaling of the threshold time $t_\text{thresh}$ 
as function of $\nx{1}$ and found that 
$t_\text{thresh} \propto \sqrt{\nx{1}}$
with a prefactor between $3$ and $4$ for $\jq=1$.
This is actually expected
for the approximation of time dependences on the basis of 
Hermite polynomials. Hence, we realize that the implementation
using the occupation number representation is not very powerful
for long times. But for the purposes of our present goal
of a proof-of-principle comparison to other data this implementation
is  sufficient.

Next, we have to include the dynamics of the Overhauser field
by increasing $\ntr$. We aim at comparisons up to $t=50/\jq$
so that it turns out that $\ntr=3$ is sufficient. The corresponding
limits for the occupation numbers 
$\nx{j}$ for $j>1$ do not need to be chosen large because the 
dynamics of the Overhauser field is governed by the rate
$\sqrt{\gamma}\jq$ which is significantly smaller. 
These numbers can be determined self-consistently by increasing
$\nx{j}$ step by step till the result no longer depends
on $\nx{j}$. In this way, we arrive at $\nx{1}=181, \nx{2}=8$,
and $\nx{3}=1$. This triple will be used henceforth, if not denoted
otherwise. 

In contrast to the solution for a
frozen Overhauser field, including the dynamics of the bath, i.e., of
the Overhauser field, leads to a further decay of the autocorrelation function $S(t)$ 
such that $S(t\to\infty) < S(t=0)/3$ holds. Describing this dynamics
of the bath quantum mechanically in the limit of very large $\neff$ is a 
central goal of the present approach.

\subsection{Other approaches}
\label{ssc:other}

We compare the data obtained from the effective Hamiltonian $\Heff$
in \eqref{eq:effect_hamil} derived from the iterated equations
of motion (iEoM) to data from three other techniques.

The first one is a fully classical simulation (classical) averaged over
random Gaussian initial configurations. Previously, it was
argued \cite{chen07} and shown that this approach approximates the quantum mechanical
dynamics fairly well \cite{stane13,stane14b} and it can
be efficiently used for large spin baths and large times
\cite{fause17a}. The spin baths studied below can easily
be treated without further approximations. The data
is averaged over $10^8$ initial configurations so that
no statistical error is discernible.

The second one is the Bethe ansatz (BA). Although it is known
since the early days of Gaudin \cite{gaudi76,gaudi83} that
the CSM is integrable and exactly solvable by Bethe ansatz,
it has taken a long time till the tedious evaluation
of the Bethe equations for larger systems has become possible
\cite{farib13a}. The treatment of the fully disordered initial
state poses an additional challenge which has been solved 
by importance sampling \cite{farib13b}. Here we use data
already computed for Ref.\ \onlinecite{seife16} to test 
rigorous bounds for persisting correlations.
The BA evaluated in the above cited fashion is very powerful
in determining the dynamics for long times, but the bath
sizes may not exceed 48 spins. Due to the stochastic
evaluation the data has a relative error of about 5\% \cite{seife16}.

The third technique is the  time-dependent density-matrix renormalization
group (DMRG). This approach is mostly used for one-dimensional
problems, but it is also perfectly suited to treat star-like
clusters as in the CSM \cite{stane13}. On the one hand, DMRG is
powerful enough to deal with up to about 1000 spins in the bath.
On the other hand, its caveat is that the growth in entanglement
is so fast that only times up to about 30 to 50$\jqm$
can be reached reliably. The parameters for the data shown
below are the following. We keep $4096$ states in the DMRG
sweeps and use the second order Trotter-Suzuki decomposition for 
the time propagation with a time step of $0.01\jqm$.
The dominant cause for the loss of accuracy is the discarded weight
in the course of the time propagation.
We stop the calculations if the accumulated discarded weight
exceeds $0.001$. 

We emphasize that the iEoM approach is tailored to capture quantum 
mechanical fluctuations of the central spin for a large number of 
effectively coupled spins $\neff$. This is
the relevant case to describe experiments on semiconductor quantum dots. 
But to gauge the introduced approach we use a rather small numbers of bath spins 
(18 to 48), which are still tractable with BA and DMRG in 
order to have exact results as reference.

\subsection{Results without external magnetic field}

First,  we focus on the CSM without external field.
The motivation is twofold. Experimentally, spin noise has become
a focus of experimental studies 
\cite{crook10,kuhlm13,zapas13a,dahba14,yang14,glase16} so that reliable 
theoretical investigations are called for. Theoretically, it
turns out that the zero-field case represents a particular challenge
because for \emph{finite} fields expansions around isolated precessing
spins work quite successfully 
\cite{khaet02,khaet03,coish04,breue04,fisch07,ferra08,deng06,deng08,coish10,barne12,beuge16,beuge17}.

\begin{figure}[htb]
	\centering
	\includegraphics[width=\columnwidth,clip]{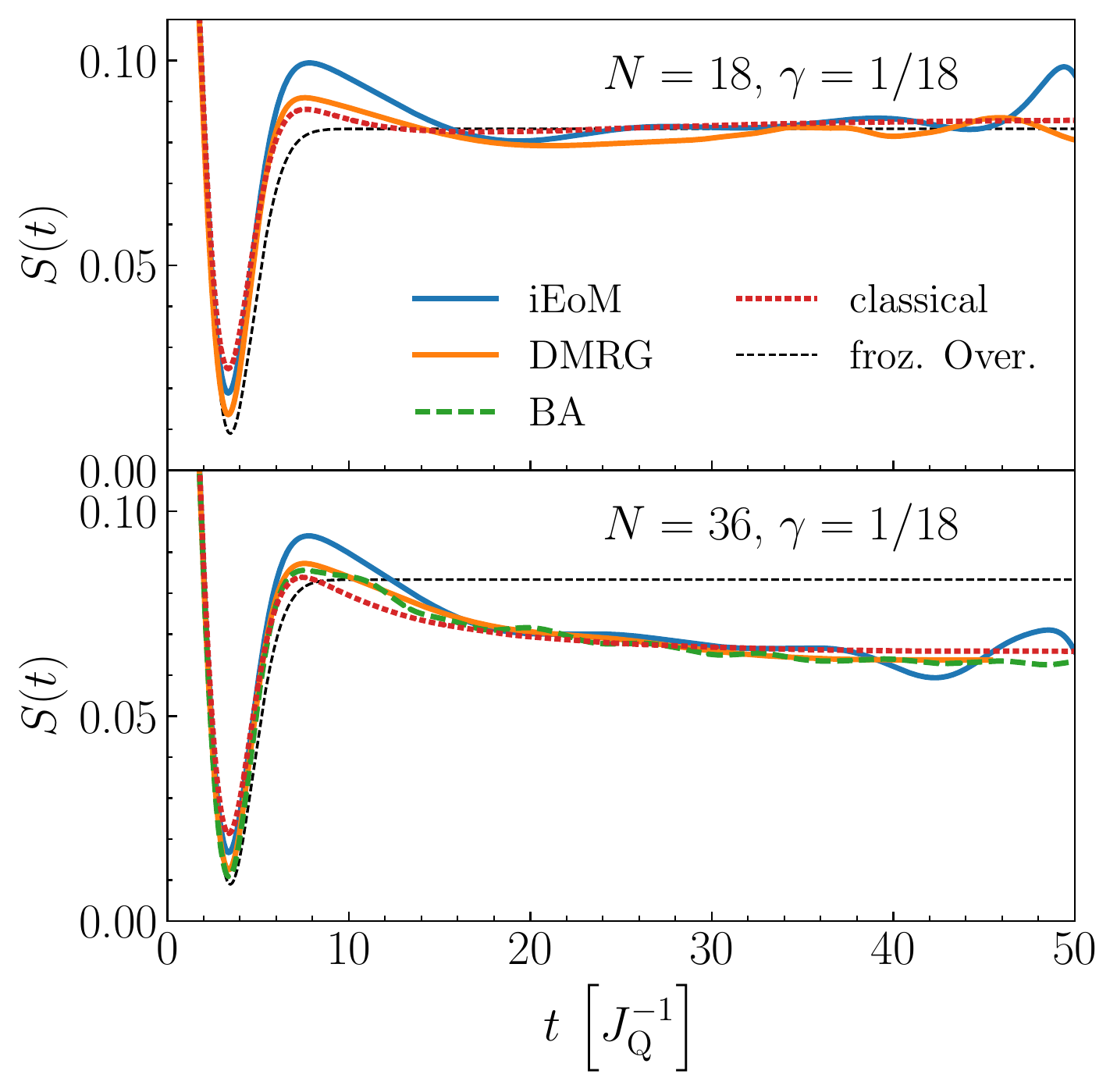}
	\caption{Comparison of results for the spin-spin correlation
	for given values of $\gamma=1/18$, cf.\ Eq.\ \eqref{eq:def_J},
	in the zero-field CSM obtained by various approaches 
	explained in the main text, see Sect.\ \ref{ssc:other}.
	The dashed line is the analytical result for frozen 
	Overhauser field \eqref{eq:fover} depicted for reference.
	The behavior for very short times is indistinguishable in all
	approaches including the frozen Overhauser field.
	The upper panel shows the curves for a smaller total number $N$ of
	spins while the lower panel refers to a larger number $N$.
	\label{fig:gamma18}}
\end{figure}

In Fig.\ \ref{fig:gamma18} we included the  static, frozen Overhauser
formula \eqref{eq:fover} for comparison with the classical, the 
fully quantum mechanical, and the iEoM approach. In the upper panel,
it appears that all approaches display the same long time behavior
close to $S(t\to\infty)=S(0)/3$. But in the lower panel, it is obvious that 
the dynamics of the bath yields a lower autocorrelation $S(t\to\infty) < S(0)/3$.
For a comprehensive rigorous discussion of this aspect we refer
the reader to Refs.\ \onlinecite{uhrig14a,seife16}, which deal with 
persisting autocorrelations in the infinite time limit in the quantum CSM.

\begin{figure}[htb]
	\centering
	\includegraphics[width=\columnwidth,clip]{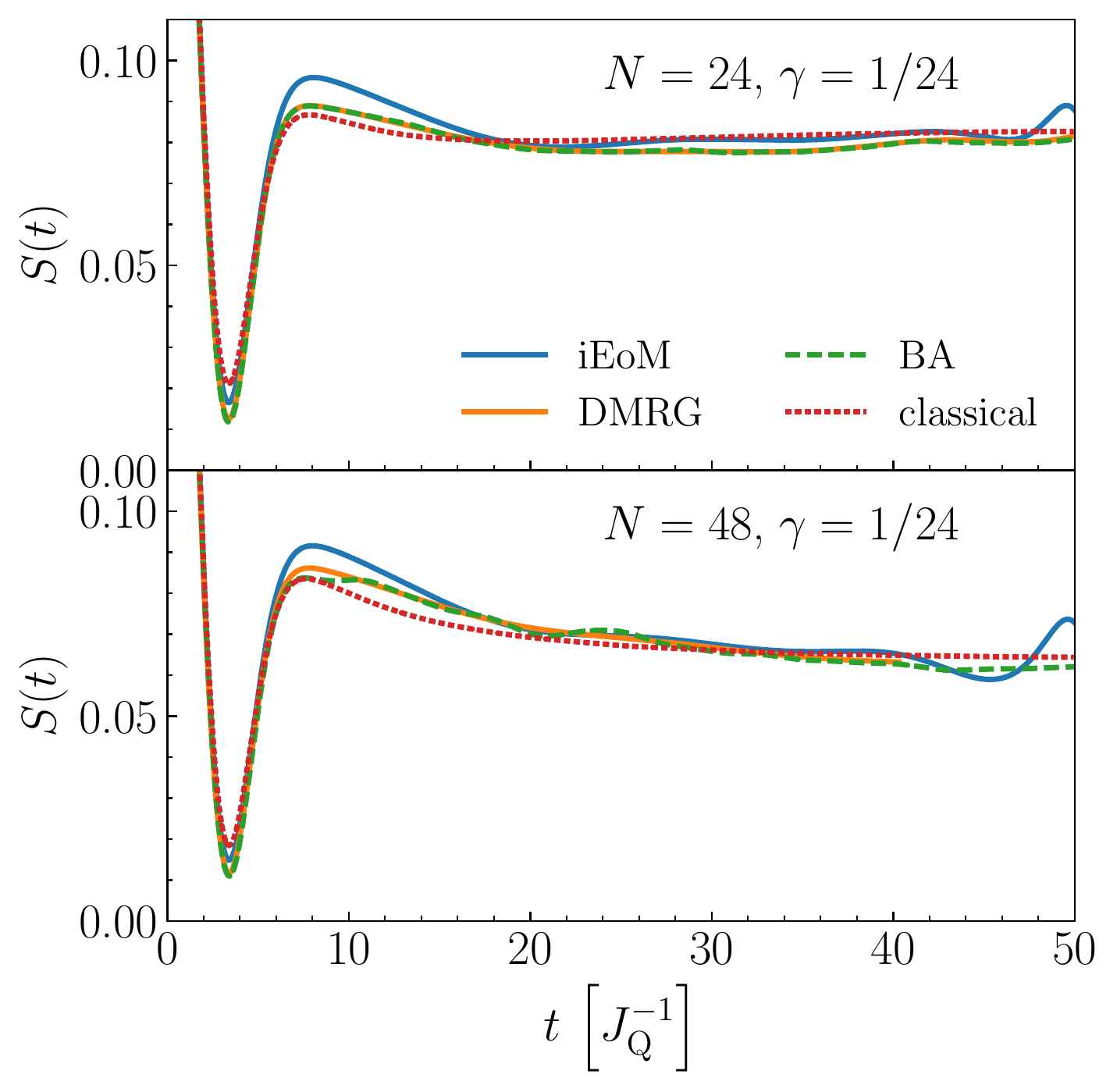}
	\caption{Comparison of results for the spin-spin correlation
	for given values of $\gamma=1/24$, cf.\ Eq.\ \eqref{eq:def_J},
	in the zero-field CSM obtained by various approaches.
	The upper panel shows the results for smaller $N$, 
	the lower panel for larger $N$.
	\label{fig:gamma24}}
\end{figure}

\begin{figure}[htb]
	\centering
	\includegraphics[width=\columnwidth,clip]{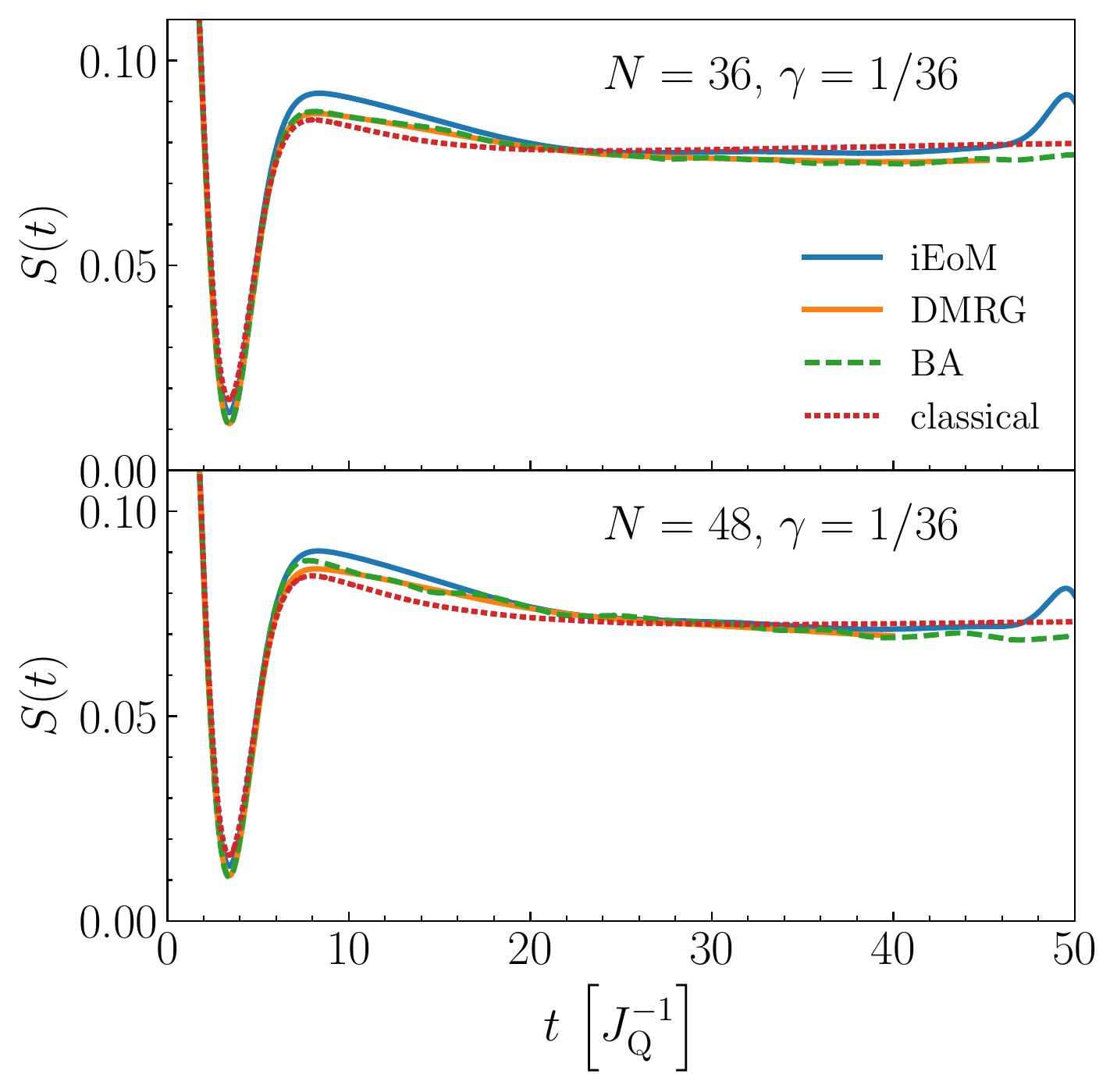}
	\caption{Comparison of results for the spin-spin correlation
	for given values of $\gamma=1/36$, cf.\ Eq.\ \eqref{eq:def_J},
	in the zero-field CSM obtained by various approaches.
	The upper panel shows the results for smaller $N$, 
	the lower panel for larger $N$.
	\label{fig:gamma36}}
\end{figure}

Figures \ref{fig:gamma18}, \ref{fig:gamma24},
and  \ref{fig:gamma36} depict a series of three decreasing
values of $\gamma$ because the derived approach (iEoM)
resides on an expansion in $1/\neff\propto\gamma$.
In each figure the upper panel shows the case of a lower value
of the total number $N$ of bath spins. The lower panel shows
data for a significantly larger $N$. At first glance, it
can be stated that all the curves agree quite well and show
the same qualitative behavior. The iEoM data shows some
wiggles starting around $t=45\jqm$ which can be attributed
to the truncation of the basis. They could be suppressed
by choosing larger $\nx{j}$, but this enhances the 
required memory exponentially and the present
calculation already used about 256 Gbytes of RAM. 
In addition, the wiggles tell us
where the truncation effects show up so that it is instructive
to see them.

\begin{figure}[htb]
	\centering
	\includegraphics[width=\columnwidth,clip]{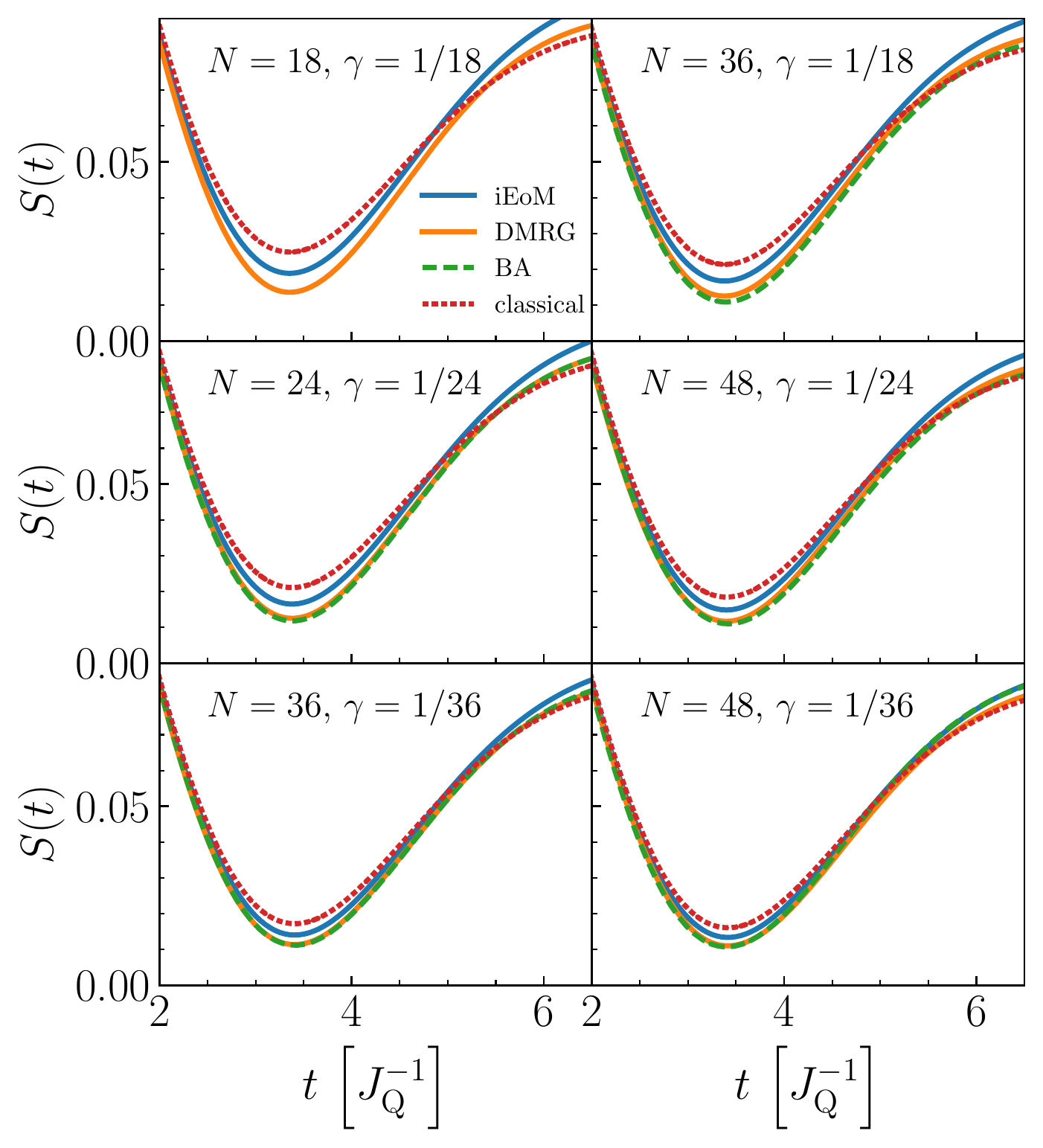}
	\caption{Comparison of results for the spin-spin correlation
	for various values of $\gamma$ 
	in the zero-field CSM obtained by various approaches.
	Here we focus on the initial dip. The left panels 
	show the results for smaller $N$, the right panels for larger $N$.
	From top to bottom, the sequence of panels 
	depicts data for decreasing values of $\gamma$.
	\label{fig:gammas}}
\end{figure}

Since the initial dynamics leading to the dips around 
$t\approx 3.5\jqm$ is very difficult to see in 
Figs.\ \ref{fig:gamma18}, \ref{fig:gamma24},
and  \ref{fig:gamma36}, we include Fig.\ \ref{fig:gammas}
where zooms of the dips are shown. From top to bottom,
the parameter $\gamma$ decreases while from left to right
the total number $N$ of spins increases.

Comparing the data of the other approaches among them
we see that the BA and the DMRG data almost coincide as it 
has to be because both approaches are numerically exact.
By this we mean that in principle, using large enough 
resources, the results can be made arbitrarily accurate.
The persisting deviations
can be attributed to statistical errors in the BA evaluation
which is based on importance sampling. They occur at shorter times
but do not accumulate for longer times.
The accuracy of the DMRG data is very high for short times,
but deteriorates for longer times for three reasons. The first
is the exponential growth of entanglement which cannot be
captured anymore by the number of states kept beyond a certain
time threshold. The second is the accumulated discarded weight
in the temporal propagation of the state. The third is the 
accumulated errors due to the Trotter-Suzuki discretization.
The main issue in the presented numerical data is discarded weight.

The classical simulation represents an approximate
treatment residing on very good arguments for the Overhauser field
for large spin baths
\cite{chen07,stane13,stane14b}, but not for the central spin.
Against this background, the proximity of the averaged classical
result to the fully quantum mechanical calculations is
remarkable.

Turning to the comparison of the iEoM results with the other data,
one becomes aware that the effects are relatively small
because all data are already close to one another.
Yet, two trends catch the eye. First, the agreement of the iEoM curves
with the DMRG and BA data becomes slowly better if at fixed $\gamma$ 
the total number $N$ of spins is increased. This effect can be
assessed by comparing the upper panel with lower $N$ to the
lower panel with larger $N$ in each of the three figures.
This is a nice observation bearing in mind that experimentally
$N$ is of the order or the number of atoms in the sample 
($N\approx10^{21}$ for a platelet  with linear dimensions
in the range of millimeters and a weight of $0.1$g), 
i.e., infinity for any practical purpose.

\begin{figure}[htb]
	\centering
	\includegraphics[width=\columnwidth,clip]{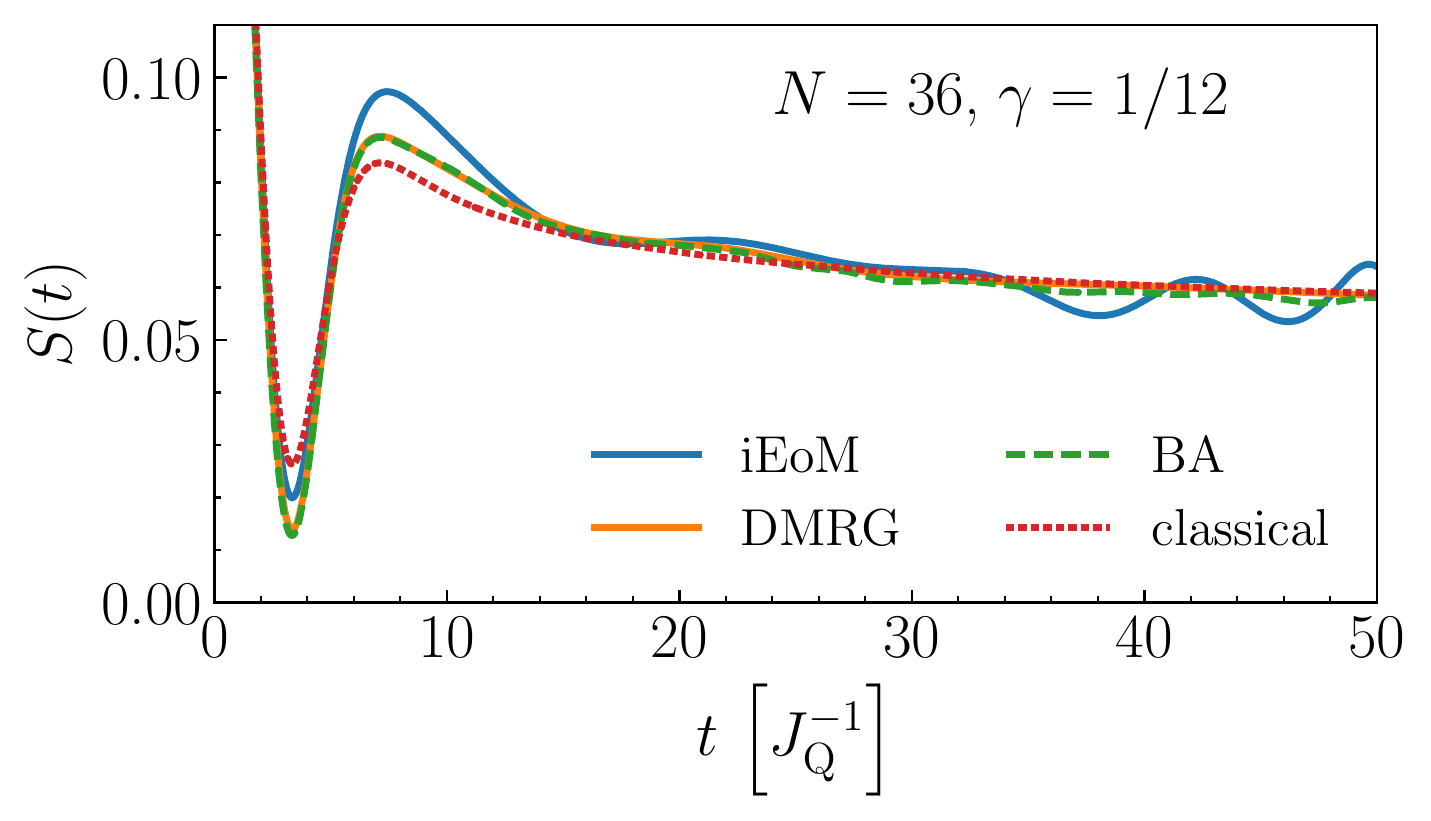}
	\caption{Comparison of results for the spin-spin correlation
	for given values of $\gamma=1/12$, cf.\ Eq.\ \eqref{eq:def_J},
	in the zero-field CSM obtained by various approaches.
		\label{fig:gamma12}}
\end{figure}

The second effect concerns the dependence on $\gamma$
and it is more pronounced. By inspecting the series of decreasing
$\gamma$ from Fig.\  \ref{fig:gamma18} via Fig.\ \ref{fig:gamma24} 
to Fig.\  \ref{fig:gamma36} or in Fig.\ \ref{fig:gammas}
from top to bottom, one notes that the agreement becomes
better and better. This was to be expected because the
approach as derived above resides on the leading
behavior in an expansion in $\gamma$. Hence, the numerical data
strongly corroborates the validity of the introduced approach.

In order to underline this aspect further, Fig.\ \ref{fig:gamma12}
depicts the result for a large value of $N$, but also a large value
of $\gamma$. This figure is to be compared to the upper panel
of Fig.\ \ref{fig:gamma36} which shows data for the same $N$, but
for a three times smaller $\gamma$. Clearly, the results 
for smaller $\gamma=1/36$ agree significantly better than the results
for $\gamma=1/12$.
We emphasize again that the experimentally relevant
 values in quantum dots are of the order of $\gamma\approx 10^{-5}$
or even smaller.
Hence, one can expect that the mapping of the CSM to an impurity in
a bosonic bath is extremely accurate and captures 
all essential physics in the CSM relevant for semiconductor quantum dots.

It is interesting that in the range around $t\approx10\jqm$ 
the deviations of the iEoM curves to the fully
quantum mechanical results are larger than for longer times 
where the agreement is much better. 
We attribute this observation to the fact that
the neglected terms in the limit  $\gamma\to0$ are of order $\gamma$
\emph{relative} to the leading order. For $\gamma=0$, only
the frozen Overhauser dynamics \eqref{eq:fover} remains
which is sizeable up to $t\approx10\jqm$, but
completely flat and featureless beyond this time.
Hence, in the time regime up to $t\approx10\jqm$ the
corrections are of order $\gamma$.
However, beyond this temporal regime,
 the dynamics is governed by $\Heff^\text{ch}$
which is of order $\sqrt{\gamma}\jq$. Hence, for longer times
the neglected corrections are of order $\gamma^{3/2}$
and thus even smaller.

Still, further support for the above promising
 conclusions would be desirable, for
instance for higher order correlations \cite{press10,becht16,frohl17} 
and for CSMs subject to pulses \cite{greil06b,greil07a,beuge16,beuge17,jasch17}.
In the present paper, we focused on the autocorrelation of the central spin
in order to establish the mapping.  But calculating other quantities such as higher order correlations is possible and cannot be done by classical means 
since the sequence of operators of the central spin matters.

\subsection{Results with external magnetic field}

In the previous  subsection we focused on the CSM without external field
for experimental and theoretical reasons.
Yet it is, of course, important to illustrate that the advocated iEoM 
approach also works with finite fields. 
So we extend the effective Hamiltonian
$\Heff$ in \eqref{eq:effect_hamil} by the dominant Zeeman term 
for the central spin in $x$-direction given
in \eqref{eq:eff-zeeman} and solve the ensuing differential equations in 
time.

\begin{figure}[htb]
	\centering
	\includegraphics[width=\columnwidth,clip]{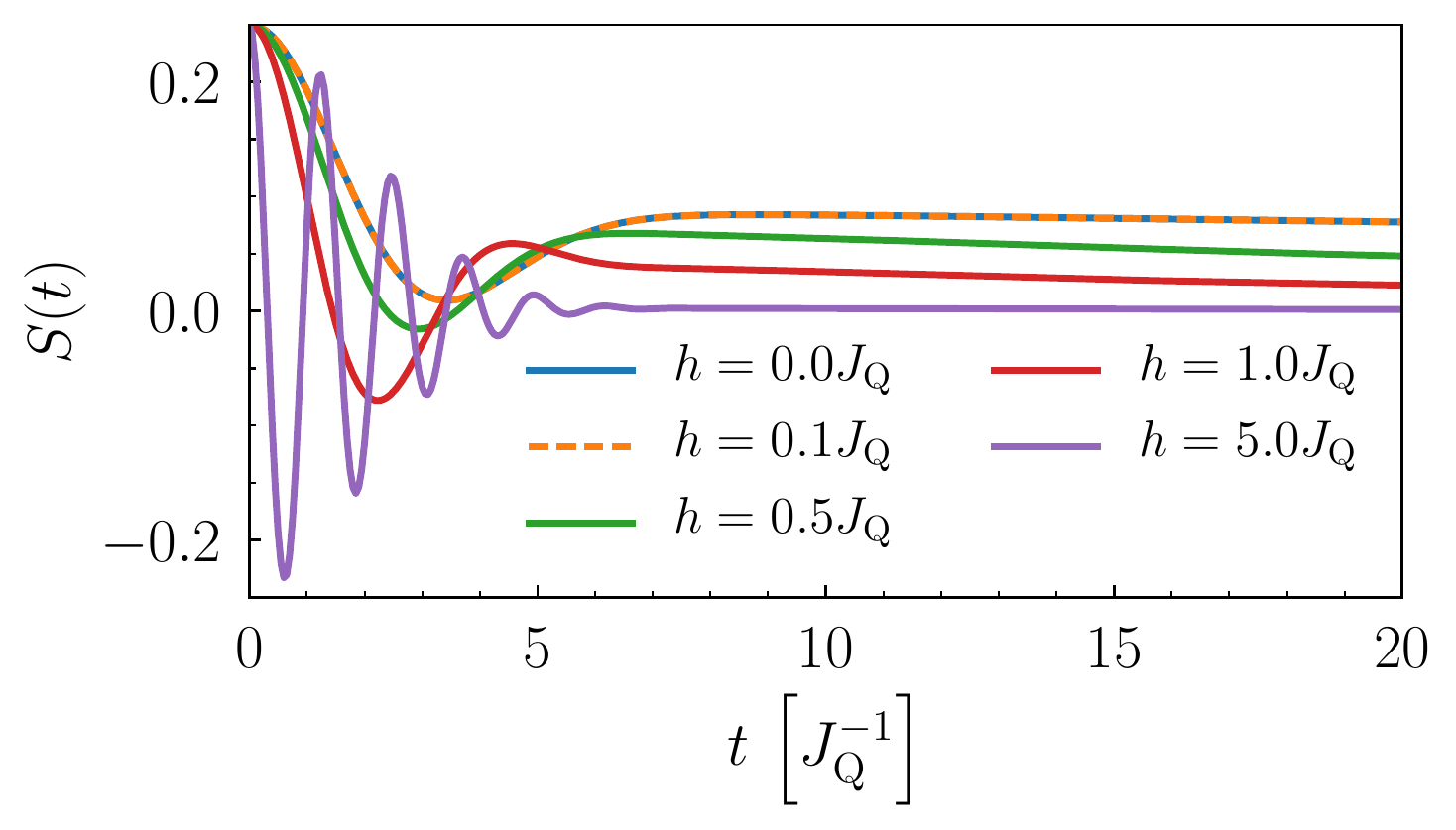}
	\caption{Spin-spin correlation from iEoM 
	for finite magnetic fields along $S^x$.
	The spin bath is infinite ($N=\infty$) with parameter $\gamma=0.01$
	corresponding to $\neff=200$ effectively coupled spins.
	The truncation of the bosonic Hilbert space is characterized
	by $\nx{1}=51, \nx{2}=2$, and zero otherwise.
		\label{fig:weak_magnet}}
\end{figure}

From numerical results \cite{hackm14a}, we expect that the Larmor
precessions are governed by a frequency which 
is altered from the case of isolated spins where it
is given by the magnetic field $h$. Due to the coupling to the
bath the energy scale $\jq$ enters yielding a shifted
Larmor frequency 
\be
\omega_\text{L} =\sqrt{h^2+\jq^2/2}.
\ee
This implies that small fields $h\ll\jq$ hardly show
any effect. Clearly, this is reproduced in Fig.\ \ref{fig:weak_magnet}
for $h=0.1\jq$. Only above $h\approx0.5\jq$ a sizable
effect sets in. True oscillations set in only above 
$h\approx5\jq$ in accordance with what was observed previously
\cite{stane14b}. For such large magnetic fields one discerns
an oscillation bounded by an envelope function. This envelope
function results from the Gaussian fluctuations of the 
Overhauser field $\ov{B}$ along the direction of the
external magnetic field. Since the Fourier transform
of a Gaussian distribution of Larmor frequencies is again a 
Gaussian in time, the envelope function is given by 
\cite{merku02,stane14b,hackm14a}
\be
S_\text{H}(t)=\frac{1}{4}\exp\left(-\frac{\jq^2}{8}t^2\right).
\label{eq:envelope}
\ee 
If the distribution of the Overhauser field is squeezed it is 
indeed possible to extend the coherence \cite{grave16}.
Alternatively, projective measurements help to maintain the central
spin polarization \cite{klaus08b}.

\begin{figure}[htb]
	\centering
	\includegraphics[width=\columnwidth,clip]{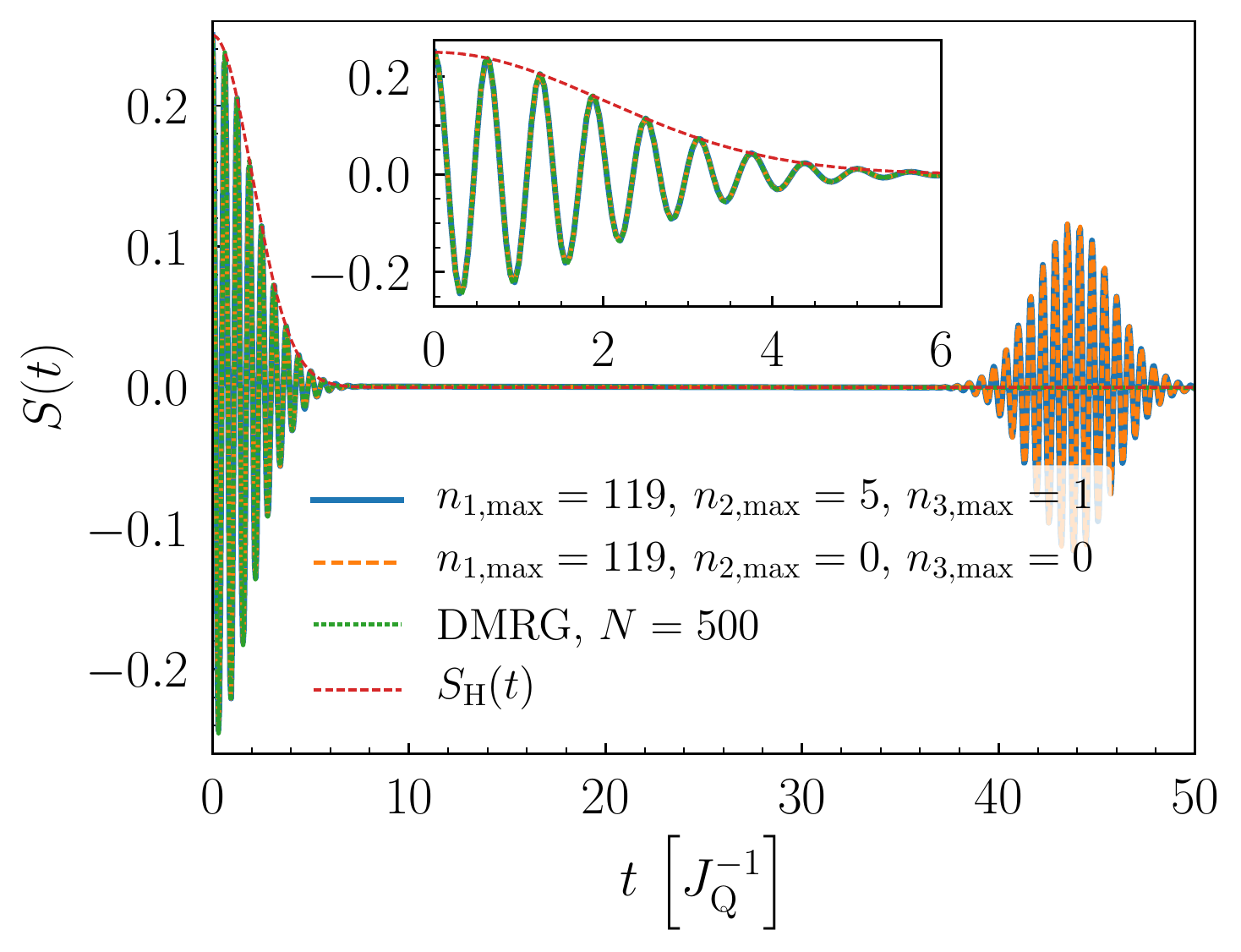}
	\caption{Spin-spin correlation from iEoM for a magnetic field $h=10\jq$
	along $S^x$. The spin bath is infinite ($N=\infty$) with parameter 
	$\gamma=0.01$ corresponding to $\neff=200$ effectively coupled spins.
	The truncation of the bosonic Hilbert space is characterized
	by the numbers given in the legend. The DMRG data shown for reference
	are obtained for the same $\gamma$ and $N=500$. The envelope 
	$S_\text{H}(t)$ is given by \eqref{eq:envelope}. The inset 
	zooms into the region of the Larmor oscillations for clarity.
		\label{fig:compar_magnet}}
\end{figure}

The excellent description of the decay of the
Larmor precession by the envelope \eqref{eq:envelope}
is illustrated in Fig.\ \ref{fig:compar_magnet}.
In addition, this figure illustrates effects of the finite
truncation of the bosonic Hilbert space. The spurious revival
can be pushed to larger times for larger $\nx{1}$. 
Note that $\nx{2}$ appears to be rather unimportant. This is so
because the signal has vanished already on short time scales so
that the dynamics of the Overhauser bath barely plays a role.

We stress that prior to the spurious revival the data obtained
by the derived iEoM agrees perfectly well with the
DMRG data included for comparison. This underlines that
the iEoM approach not only works nicely in the zero-field
case, but also at finite magnetic fields. 
In contrast to DMRG, the iEoM approach allows one to
tackle much larger values of $\neff\propto 1/\gamma$ in the
limit $N=\infty$.

\section{Conclusions and Outlook}
\label{sec:conclusion}

Motivated by the importance of systems of a single spin
coupled to large baths of spins in quantum dots 
\cite{merku02,schli03,hanso07,coish09,fisch09,urbas13,ribei13,warbu13,chekh13b}, 
but also in NV-centers in diamond \cite{jelez06} or in generic NMR studies
\cite{alvar10c}, we investigated the quantum mechanical 
central spin model in the limit
of large numbers $\neff$ of bath spins.
In quantum dots, $\neff$ can be as large as $10^5$ to $10^6$ and
around NV-centers and in large organic molecules the number
of effectively coupled spins still ranges from 10 to 100.

Theoretically, we posed ourselves the question
whether there is a well-defined limit $\neff\to\infty$
and if so, whether the system becomes classical
in this limit \cite{chen07,stane13,stane14b}.

We started from the equations of motion for the spin operators,
employed a suitable scalar product for operators \cite{kalth17}, and used
the observation that traces over infinite sums of spin operators
can be computed from classical Gaussian correlations \cite{seife16}.
In this way, we mapped the operator dynamics in the 
central spin model in the limit $\neff\to\infty$
to the dynamics of states in an effective quantum model
with a four-dimensional central impurity coupled to 
a bosonic bath without further interaction.
The four-dimensional impurity represents the possible
operators for the central spin. The bosons in the bosonic bath
represent the collective spin degress of freedom in 
the large spin bath.
Hence, a well-defined limit $\neff\to\infty$ has been established.

We find it very remarkable that the analytic treatment
of the limit $\neff\to\infty$
does not make the system completely classical, but
keeps its quantumness in the effective Hamiltonian $\Heff$. 
Yet, the numerical data obtained from the averaged classical
calculation, from the numerically exact approaches, and from the iEoM
approach are very close to one another.
The treatment of external magnetic fields
is also possible. This holds for the central spin, but also
for the relevant nuclear Zeeman terms \cite{beuge17,jasch17}.

The derived mapping has two fundamental advantages: (i)
It paves the way to the quantum mechanical treatment of
very large spin baths which cannot be tackled otherwise 
at all. (ii) It enables the treatment of the central spin
model by techniques which could so far not be used, for instance
any approach conceived to deal with impurities coupled to
interaction-free bosonic baths.

We tested the results obtained by the iEoM approach 
for smaller spin baths with 18 to 48 spins where reliable results
obtained by established techniques are available. In this way,
we verified the validity of the analytic arguments used
in the derivation. The results are remarkably close
to the numerically exact reference data, although the relatively low numbers 
of bath spins are disadvantageous for the introduced approach
to work perfectly.
The agreement improved for larger spin baths corroborating the 
derivation based on an expansion in $1/\neff$. 

Moreover, the good agreement for baths
of moderate size suggests that the application of the 
iEoM approach is already fruitful for smaller spin baths
as they arise for NV centers or in large molecules.

In cases of larger spin baths, the approach is expected to
yield highly accurate results where other methods can not be applied at all. 
Thus the iEoM allows for investigations of quantum effects beyond classical
descriptions in the regime of large spin baths. It
has been the main goal of the present paper to derive 
such a theoretical technique.

Furthermore, the present treatment can
be extended in a straightforward manner to larger bath spins,
for instance $S=3/2$, which is the relevant case in GaAs \cite{beuge17}.
To this end, only the variance of the Gaussian distributions
has to be adapted.

We point out that the numerical implementation
of the mapped model is not yet pushed to its limits. Further improvements
are called for in order to reach longer times. This is crucial
to describe many experiments, for example for measurements of
 higher order correlations \cite{press10,becht16,frohl17} or  
for quantum dots prepared in non-equilibrium
states by intricate pulsing 
\cite{greil06b,greil07a,beuge16,beuge17,jasch17}.

The route to follow to reach the necessary improvement
is to exploit the significant difference in dynamics
induced by the Hamiltonian part $\Heff^\text{CS}$
and by the Hamiltonian part $\Heff^\text{ch}$ where
the rate of changes induced by the latter is smaller
by a factor $1/\sqrt{\neff}$. Hence, one should treat 
$\Heff^\text{CS}$ exactly by choosing a representation
in which it is diagonal. Such implementations are left for
future research.

\begin{acknowledgments}
We thank Frithjof B.\ Anders, Philip Bleicker, and Mohsen Yarmohammadi
for helpful discussions. We are indebted to Alexandre Faribault for 
provision of data. This study has been
supported financially by the Deutsche Forschungsgemeinschaft (DFG) 
and the Russian Foundation for Basic Research in International
Collaborative Research Centre 160 and by the DFG in grant UH 90/9-1.
\end{acknowledgments}

%\bibliographystyle{apsrev}
%\bibliography{../../bibinput/liter10}

\begin{appendix}

\section{Orthonormality of the generalized Overhauser fields}
\label{app:ortho}

The fact that we orthonormalized the polynomials according to 
\eqref{eq:onb1}
allows us to conclude that the components of the generalized Overhauser fields
$\ov{P}_j$ are pairwise orthonormal with respect to the Frobenius scalar product
\eqref{eq:def_scalar} as well. For $j,k\in\{1,2,\ldots,\ntr\}$ and $\alpha,\beta
\in\{x,y,z\}$ we obtain
\begin{subequations}
\label{eq:orthonormal}
	\begin{align}
		&\lgl{P}_{j}^{\alpha}\left|{P}_{k}^{\beta}\right.\rgl
		= \nonumber \\
		&\quad =		\frac{1}{2^{N-1}}
		\tr\left(\sum_{n,m=1}^{N}p_j(J_n)p_k(J_m)S_{n}^{\alpha} S_{m}^{\beta}\right)
		\\&\quad=
		\frac{1}{2^{N-1}}
		\sum_{n,m=1}^{N}p_j(J_n)p_k(J_m)
		\underbrace{\tr\left(S_{n}^{\alpha}S_{m}^{\beta}\right)}_{2^{N-1}\delta_{nm}\delta_{\alpha\beta}}
		\\&\quad=
		\sum_{n=1}^N p_j(J_n)p_k(J_n)\delta_{\alpha\beta}
		\\&\quad=\delta_{jk}\delta_{\alpha\beta} 
	\end{align}
\end{subequations}
which was to be proven.

\section{Irrelevance of commutators in the limit $\neff\to\infty$}
\label{app:commute}

Here we provide an example that commutators represent subleading
corrections to the scalar products of the products of
generalized Overhauser fields. In order to keep the example
transparent we consider the simplest commutator of
two linear fields
\bs
\begin{align}
	[P^\alpha_m,P^\beta_l]
	&= 	\sum_{k,j=1}^{N}
	p_m(J_k)p_l(J_j) 4 [S^\alpha_k,S^\beta_j]
	\\&=
	\sumin	4p_m(J_k) p_l(J_k) i\sum_\delta \epsilon_{\alpha\beta\delta} S^\delta_k
	\\&=:I^\delta.
\end{align}	
\es
Next, we consider the norms of the involved operators.
The generalized Overhauser fields $P^\alpha_k$, $P^\beta_l$ are normalized
to unity by construction. But the scalar product of $I^\delta$ is of order $1/\neff$
because each factor  $p_m(J_k)$ and  $p_l(J_i)$ is of order
$1/\sqrt{\neff}$. Hence, the summation over all bath spins in
$\lgl I^\delta|I^\delta\rgl$  sums $\neff$ terms, each of which proportional
to $p_m(J_k)^2 p_l(J_k)^2$ and hence of order $1/\neff^2$.
So, the scalar product of $I_\delta$ is of order $1/\neff$ and hence 
subleading relative to the contributions resulting from products of the
generalized Overhauser fields. 

The above example illustrates the
statement from the main text that the sequence of the generalized Overhauser
fields in the terms of the operator basis does not matter
for $\neff\to\infty$.

\section{Calculation of $T_1$ and $T_2$}
\label{app:t_1_t_2}

In order to obtain $T_1$ in Eq.\ \eqref{eq:t1}, we compute
\begin{align}
& \Li\sigma_m = 
		\label{eq:kommutatorI}
\\
\nonumber
& \left\{
\begin{array}{ll}
		0 & \text{if}\;  m=0 
		\\
		\frac{i}{2}\s{m+1}{}\P{1}{m-1}-\frac{i}{2}\s{m-1}{}\P{1}{m+1} & 
		\text{if}\; m\in\{1,\,2,\,3\},
		\end{array} \right.
\end{align}
where the indices for finite $m$ must be understood in a cyclic sense, i.e., 
for $m=1$ the decremented $m-1$ means $m-1=3$ and for 
$m=3$ the incremented $m+1$ means $m+1=1$.
If we use the result \eqref{eq:kommutatorI} to compute the term $T_1$
for $\Li(\sigma_m A^{\m{n}})$ we obtain
\begin{align}
&T_1[\Li(\sigma_m A^{\m{n}})] =
\nonumber \\
&\quad \frac{i}{4}
\left(\sigma_{m+1}\{\P{1}{m-1},A^\m{n}\} -\sigma_{m-1}\{\P{1}{m+1},A^\m{n}\} 
\right).
\end{align}
The multiplication with $\P{1}{m\pm1}$ modifies the index of the Hermite polynomials 
 $H_n(\P{1}{m\pm1})$ according to \eqref{eq:hermite_argument}. 
This can be concisely expressed by creation and annihilation operators
as is common in the analytic treatment of the eigen wave functions of the
harmonic oscillator.

In order to address $T_2$ in Eq.\ \eqref{eq:t2}, we consider terms of the type
\begin{align}
	\Li\Hn{j}{\alpha}(\P{j}{\alpha}) = [\Hh,\Hn{j}{\alpha}(\P{j}{\alpha})].
	\label{eqn:kommutator_hermite}
\end{align}
First, we compute the application of the Liouville operator to the component of 
a generalized Overhauser field $\P{j}{\alpha}$. 
A straightforward calculations shows that \eqref{eq:lanczos} implies
\begin{align}
	\Li\ov{P}_j =
	-i
	\ov{S}_0\times
	\left(
		\beta_j\ov{P}_{j+1}+\alpha_j\ov{P}_j+\beta_{j-1}\ov{P}_{j-1}
	\right)
	\label{eq:liouville_feld}
\end{align}
with the coefficients $\alpha_j$ and $\beta_j$. The coefficients 
$\beta_0:=0$ and $\beta_{\ntr}:=0$ are defined to vanish. 
If we use the definition of $R_j^\delta$ in \eqref{eq:def_R}
we can express the outer product in \eqref{eq:liouville_feld}
using the Levi-Civita symbol $\levi{\alpha}{\beta}{\delta}$
\begin{align}
\label{eq:kreuzprodukt}
	\Li\P{j}{\alpha} =
	-i \sum_{\beta,\delta=1}^3 \levi{\alpha}{\beta}{\delta}
	S_0^\beta R_j^\delta.
\end{align}

In order to treat the general case in \eqref{eqn:kommutator_hermite}
we combine \eqref{eq:kreuzprodukt} with the operator relation
\begin{align}
\label{eqn:kommutator_diff}
	[A,f(B)] &= \frac{\partial f(B)}{\partial B} [A,B].
\end{align}
The latter requires that $[B, [A,B]]=0$  which does not hold generally.
But here we can safely assume commutativity because
the sequence of bath operators does not matter in leading order in $1/\neff$. 
In this way, we arrive at
\bs
	\begin{align}
	[\Hh,\Hn{j}{\alpha}] &=-i\Hn{j}{\alpha}^\prime
	\sum_{\beta,\delta=1}^3 \levi{\alpha}{\beta}{\delta}
	S_0^\beta R_j^\delta
		\\&=
	-i\sqrt{n_{j,\alpha}}H_{n_{j,\alpha}-1} 
	\sum_{\beta,\delta=1}^3 \levi{\alpha}{\beta}{\delta}
	S_0^\beta R_j^\delta.
	\label{eqn:kommutator_hermite_voll}
	\end{align}
	%\label{eqn:hermite-kommutator_voll_ganz}	
\es
Hence, we have to decrement the degree of the 
Hermite polynomial $H_{n_{j,\alpha}}$. 

All other Hermite polynomials are not affected in this step and
remain unaltered. This is precisely what 
 the decrementing mapping $\mpi(\m{n},j,\alpha)$
defined in Eq.\ \eqref{eq:mpi}  expresses.
In this way, we arrive at the final result for the product
of Hermite polynomials as it appears in $A_\m{n}$
\begin{align}
	[\Hh,A_\m{n}] &=
	-i\sum_{j=1}^{\ntr}\sum_{\alpha=1}^3
	\sqrt{n_{j,\alpha}}A_{\mpi(\m{n},j,\alpha)}
	\sum_{\beta,\delta=1}^3 \levi{\alpha}{\beta}{\delta}
	S_0^\beta R_j^\delta.
\end{align}
Finally, we combine this finding with the Pauli matrices for
the central spin and obtain
\bs
\be
T_2[\Li(\sigma_m A_{\m{n}})] =  \frac{-i\sigma_0}{2} \sum_{j=1}^{\ntr}
		\sum_{\alpha,\delta=1}^3 \levi{\alpha}{m}{\delta} \sqrt{n_{j,\alpha}}
		R^\delta_j A_{\mpi(\m{n},j,\alpha)}
\ee
for $m\in\{1,2,3\}$ where we used that 
$\{\sigma_m,\sigma_\beta\}=2\delta_{m\beta}$
if $m,\beta\in\{1,2,3\}$. For $m=0$, we obtain
\be
T_2[\Li(\sigma_0 A^{\m{n}})] =  \frac{-i}{2} \sum_{j=1}^{\ntr}
		\sum_{\alpha,\beta,\delta=1}^3\levi{\alpha}{\beta}{\delta} 
		\sqrt{n_{j,\alpha}} \sigma_\beta R^\delta_j	A_{\mpi(\m{n},j,\alpha)},
\ee
\es
where we used $\{\sigma_0,\sigma_\beta\}=2\sigma_{\beta}$.

\section{Commutation and Anticommutation Matrices}
\label{app:matrices}

The anticommutation of the operators of the central spin with $\sigma_k$
is described by the matrices $M_k$. Their matrix elements
are defined in Eq.\ \eqref{eq:pauli_anticommut}. The matrices read
\bs
\begin{align}
{M}_1&=
\begin{pmatrix}
0 & 1 & 0 & 0 \\
1 & 0 & 0 & 0 \\
0 & 0 & 0 & 0 \\
0 & 0 & 0 & 0
\end{pmatrix}
,
\\
{M}_2&=
\begin{pmatrix}
0 & 0 & 1 & 0 \\
0 & 0 & 0 & 0 \\
1 & 0 & 0 & 0 \\
0 & 0 & 0 & 0
\end{pmatrix} 
,
\\
{M}_3&=
\begin{pmatrix}
0 & 0 & 0 & 1 \\
0 & 0 & 0 & 0 \\
0 & 0 & 0 & 0 \\
1 & 0 & 0 & 0
\end{pmatrix}.
\end{align}
\es

The commutation of the operators of the central spin with $\sigma_k$
is described by the matrices $K_k$. Their matrix elements
are defined in Eq.\ \eqref{eq:pauli_commut}. The matrices read
\begin{subequations}
	\begin{align}
	{K}_1 &=
	\begin{pmatrix}
	0 & 0 & 0 & 0 \\
	0 & 0 & 0 & 0 \\
	0 & 0 & 0 & i \\
	0 & 0 & -i & 0
	\end{pmatrix},\\
	{K}_2&=
	\begin{pmatrix}
	0 & 0 & 0 & 0 \\
	0 & 0 & 0 & -i \\
	0 & 0 & 0 & 0 \\
	0 & i & 0 & 0
	\end{pmatrix},\\
	{K}_3&=
	\begin{pmatrix}
	0 & 0 & 0 & 0 \\
	0 & 0 & i & 0 \\
	0 & -i & 0 & 0 \\
	0 & 0 & 0 & 0
	\end{pmatrix}.
	\end{align}	
\end{subequations}

\section{Diagonalization of $\Heff^\text{ch}$}
\label{app:chain-diag}

Instead of considering a semi-infinite chain with hopping
bosons as described by $\Heff^\text{ch}$ in Eq.\ \eqref{eq:chain-hamil},
it can be advantageous to restrict it to local processes only.
Then the system resembles a bath of \emph{bosons} with flavor
$x,y,z$ connected to the central four-dimensional impurity.
This mapping is obtained by defining vectors of 
the bosonic operators
\bs
	\begin{align}
	\m{a}_\alpha^{\phantom\dagger} &:= 
	\begin{pmatrix}
	{a}_{1,\alpha}^{\phantom\dagger}, & {a}_{2,\alpha}^{\phantom\dagger}, 
	& \dots, & {a}_{L,\alpha}^{\phantom\dagger} \\
	\end{pmatrix}^\text{T},
	\\
	\m{a}_\alpha^\dagger &:= 
	\begin{pmatrix}
	{a}_{1,\alpha}^\dagger, & {a}_{2,\alpha}^\dagger, & \dots, 
	& {a}_{L,\alpha}^\dagger
	\end{pmatrix},
	\end{align}	
\es
where $\m{a}_\alpha^{\phantom\dagger}$ 
is a column vector while $\m{a}_\alpha^\dagger$
is a row vector. With these definitions and the tridiagonal
matrix $\m{T}$ from \eqref{eq:tridiagonal} 
the chain Hamilton can be denoted concisely
\begin{align}
\Heff^{\text{ch}}=\frac{i}{2}\sum_{\alpha,\beta,\delta}
\levi{\alpha}{\beta}{\delta}
{M}_\beta\ \m{a}_\delta^{\dagger}
\m{T}
\m{a}_\alpha^{\phantom{\dagger}}.
\end{align}

The tridiagonal matrix  $\m{T}$ is real and symmetric so that
it can be diagonalized by a real orthogonal matrix $\m{Q}$
\begin{align}
\m{D}=\m{Q}^{\dagger}\m{T}\m{Q}
\end{align}
yielding the diagonal matrix $\m{D}$ which has the
eigen values $\varepsilon_j$ on its diagonal. Then the transformed
vectors of  annihilation and creation operators 
$d_{j,\alpha}^{\phantom{\dag}}$ and $d_{j,\alpha}^\dag$ are given by
\bs
\label{eq:transform}
	\begin{align}
	\m{d}_\alpha
	&:= 
	\m{Q}^{\dagger}\m{a}_\alpha^{\phantom\dagger},
	\\
\m{d}_\alpha^{\dagger} 
	&:= 
	\m{a}_\alpha^{\dagger}\m{Q}.
	\end{align}	
\es
They allow us to express the Hamiltonian $\Heff^{\text{ch}}$
in an (almost) diagonal form
\begin{align}
\label{eq:chain-diag}
\Heff^{\text{ch}} =\frac{i}{2}\sum_{\alpha,\beta,\delta}
\levi{\alpha}{\beta}{\delta}
{M}_\beta \sum_{j=1}^{\ntr}
\varepsilon_j
\
{d}_{j,\delta}^{\dagger}
{d}_{j,\alpha}^{\phantom{\dagger}}.
\end{align}	

The missing piece is the transformed Hamiltonian 
$\Heff^\text{CS}$ which is obtained by inserting 
the inverses of Eqs.\ \eqref{eq:transform} into \eqref{eq:HCS}
\be
\label{eq:CS-diag}
\Heff^\text{CS} = \frac{1}{2}
\sum_\alpha  K_\alpha \sum_{j=1}^{\ntr} 
Q_{1,j}\left(d_{j,\alpha}^{\phantom{\dagger}}
+ d_{j,\alpha}^\dagger \right),
\ee
where we used the fact that $\m{Q}$ is real so that
no complex conjugation needs to be taken into account.
For simplicity, one may make all $Q_{1,j}\ge 0$ by
choosing the signs of the $d_{j,\alpha}$ appropriately.
We highlight this fact because it allows us to 
point out the relation to the spectral density approach
introduced in  Ref.\ \onlinecite{fause17a}. The exponential
discretization of the spectral densities relevant for the
set of couplings $J_i$ directly provides energies $\varepsilon_j$
and weights $W_j$. The square roots of the weights determine
the coefficients $Q_{1,j}=\sqrt{W_j}$.

Eventually, Eqs.\ \eqref{eq:chain-diag} and \eqref{eq:CS-diag} 
together define the complete effective Hamiltonian
for the central spin model with large spin baths.
This effective model comprises a central four-dimensional impurity
coupled to a surrounding bath of bosonic sites $j$, where bosons of
three flavors are exchanged with one another.

\end{appendix}

\end{document}